%% file: ergo_param.tex
\documentclass[twocolumn,superscriptaddress,showpacs,prd,aps,amsmath,amssymb,nofootinbib]{revtex4-1}                                        
\usepackage{natbib}
\usepackage{graphicx,color}
\usepackage{amsmath,amssymb}
\usepackage{verbatim}
\usepackage{float}
\usepackage{wasysym}
\usepackage{amssymb,graphicx}
\usepackage{epsfig}
\usepackage{psfrag}
\usepackage{dsfont}
\usepackage{amsfonts}
\usepackage{mathrsfs}
\usepackage{multirow}
\usepackage{times}
\usepackage{bm}
\usepackage{hyperref}
\usepackage{xspace}
\hypersetup{
  colorlinks=true,        
  linkcolor=blue,         
  citecolor=cyan,         
}
\usepackage{pifont}
\usepackage[normalem]{ulem}   
\usepackage{grffile}

\graphicspath{{./}}

\input kigou.tex

\newcommand{\GA}{\alpha}

\newcommand{\GG}{\gamma}

\newcommand{\GE}{\epsilon}
\newcommand{\GK}{\kappa}

\newcommand{\GR}{\rho}
\newcommand{\GS}{\sigma}
\newcommand{\GT}{\tau}

\newcommand{\GO}{\omega}

\newcommand{\GP}{\phi}

\newcommand{\GU}{\theta}

\newcommand{\be}{\begin{equation}}
\newcommand{\ee}{\end{equation}}

\newcommand{\Mtov}{$M_{\rm max}^{\scalebox{0.7}{TOV}}$}
\newcommand{\Mkep}{$M_{\rm max}^{\scalebox{0.7}{Kep}}$}
\newcommand{\rhotov}{$\rho_{0\rm max}^{\scalebox{0.7}{TOV}}$}
\newcommand{\rhokep}{$\rho_{0\rm max}^{\scalebox{0.7}{Kep}}$}

\def\QEQ{{%
    \setbox0\hbox{$I$}%
    \rlap{\hbox to \wd0{\hss--\hss}}\box0
}}

\begin{document}

\title{ Locating ergostar models in parameter space   }

\author{Antonios Tsokaros}
\affiliation{Department of Physics, University of Illinois at Urbana-Champaign, Urbana, Illinois 61801, USA}
\email{tsokaros@illinois.edu}

\author{Milton Ruiz}
\affiliation{Department of Physics, University of Illinois at Urbana-Champaign, Urbana, Illinois 61801, USA}
\author{Stuart L. Shapiro}
\affiliation{Department of Physics, University of Illinois at Urbana-Champaign, Urbana, Illinois 61801, USA}
\affiliation{Department of Astronomy and NCSA, University of Illinois at Urbana-Champaign, Urbana, Illinois 61801, USA}

\date{\today}

\begin{abstract}
Recently, we have shown that dynamically stable ergostar solutions (equilibrium neutron
stars that contain an ergoregion) with a compressible and causal equation of 
state exist
[A. Tsokaros, M. Ruiz, L. Sun, S. L. Shapiro, and K. Ury{\=u}, Phys. Rev. Lett. {\bf 123}, 231103 (2019)].
These stars are hypermassive, differentially 
rotating, and highly compact. In this work, we make a systematic study of equilibrium models
in order to locate the position of ergostars in parameter space. We adopt four
equations of state that differ in the matching density of a maximally stiff core.
By constructing a large number of models both with uniform and differential rotation
of different degrees, we identify the parameters for which ergostars appear.
We find that the most favorable conditions for the appearance of dynamically stable ergostars are
a significant finite density close to the surface of the star (i.e.,
similar to self-bound quark stars) and a small degree of differential rotation.
\end{abstract}

\maketitle

\section{Introduction}

One important open question in modern astrophysics is the mechanism that
powers relativistic jets in short gamma-ray bursts like the one 
accompanying event GW170817 
\cite{TheLIGOScientific:2017qsa,Savchenko:2017ffs,Savchenko17GCN}. 
Within the membrane paradigm, these highly energetic phenomena are typically 
attributed to the black hole horizon \cite{Thorne86}. On the other hand, according 
to more recent studies \cite{Komissarov:2004ms,2005MNRAS.359..801K, Ruiz:2012te}, 
it is the ergosphere
and its threading by magnetic field lines that is chiefly responsible for
the jet's existence, while a black hole horizon is not necessary.
An ergostar is a star that contains an ergoregion, i.e., a region where 
there are no timelike static observers and all trajectories (timelike or
null) must rotate in the direction of rotation of the star (frame dragging).
For an ergostar, the question relevant  to jet formation is whether 
the ergoregion is preserved in the presence of a dissipative mechanism, such as 
viscosity or a turbulent magnetic field. In particular, when an ergostar is threaded by a magnetic 
field, is stability maintained over many Alfv\'en timescales, or does the turbulent magnetic 
viscosity destabilize the star before a jet can be launched?

The question regarding the dynamical stability of an ergostar with a causal and compressible
equation of state (EOS) was answered positively in \cite{Tsokaros:2019mlz}. There, the
ALF2cc EOS was adopted to create ergostars that evolved stably for $\sim 150$
dynamical times or $\sim 30$ rotation periods. These equilibria, when perturbed
in a radial or nonaxisymmetric way, showed no significant mode growth, while their shape 
and ergoregion remained intact. At the same time, polytropic models of ergostars presented in
\cite{1989MNRAS.239..153K} proved to be unstable to radial collapse. The secular 
evolution of stars containing ergoregions is governed by the 
fact that the timelike Killing vector associated with the stationarity of the 
spacetime becomes spacelike inside the ergoregion, which implies a negative energy
with respect to an asymptotic observer for a freely moving particle there. As a 
consequence, a nonaxisymmetric perturbation that radiates positive energy at infinity
will make the negative energy in the ergoregion even more negative, leading to the 
so-called Friedman instability \cite{1978CMaPh..63..243F, 2018CMaPh.358..437M}. 
The timescale for this instability was initially considered to be longer than the
Hubble time \cite{Comins1978}, but more recently, it was found that it can be quite 
small \cite{1996MNRAS.282..580Y, Brito:2015oca} (for small mode numbers).

On the other hand, since the original work of Wilson \cite{1972ApJ...176..195W}, it seems
that differential rotation plays a crucial role in the appearance of an ergoregion
(see also the models of \cite{1975ApJ...200L.103B,1989MNRAS.239..153K, Ansorg:2001pe}).
In the presence of magnetic fields, this differential rotation is eventually
suppressed due to magnetic winding and the magnetorotational instability
\cite{Shapiro:2000zh,dlsss06a}, 
which in turn implies that jet formation (if dependent on the 
existence of the ergoregion) may be inhibited.
Given the fact that we were able to construct dynamically stable, differentially rotating 
ergostars, the answer regarding the ergosphere hypothesis on jet formation depends 
crucially on whether a dissipative mechanism will affect the structure of the ergostar 
sufficiently to remove the ergoregion before powering a jet. 
Alternatively, they may drive an ergostar, if hypermassive, to collapse to a black hole. If the ergosphere 
is indeed responsible for the formation of a jet, then in this case, its lifetime may be
different from the case where a black hole is the power source.

In order to probe the possible scenarios described above, one needs to know
the most favorable equilibria that contain ergoregions. Is it possible to have 
ergostars that are uniformly rotating (with a compressible and causal EOS)? Is it possible
to have supramassive \cite{1992ApJ...398..203C} ergostars, i.e., 
uniformly rotating ergostars with mass larger 
than the maximum Tolman-Oppenheimer-Volkoff (TOV) limit, but less than the maximum mass at 
the mass-shedding (Kepler) limit? To answer such questions,
we perform in this work a parameter study probing the existence of ergostars. Using 
four EOSs and different degrees of differential rotation, we map the location of the
ergostars on mass vs central density diagrams. Our EOS strategy is similar to 
the one employed in \cite{Tsokaros:2019mlz}. This time, we start with the SLy
EOS \cite{Douchin01} and construct a large number of uniformly and differentially
rotating models using 5 degrees of differential rotation. Then we construct three
additional EOSs based on the SLy one where we progressively substitute an inner core
at matching densities $\GR_{0\rm nuc}, 2\GR_{0\rm nuc}, 4\GR_{0\rm nuc}$ with the 
maximally stiff EOS, which has the speed of sound equal to the speed of light 
[see Eq. (\ref{eq:eoscc})]. Here,
$\GR_{0\rm nuc}=2.7\times 10^{14}\ {\rm g/cm^3}$ is nuclear matter density. 
Sequences of constant angular momentum and constant rest mass are constructed and 
stability questions are addressed.

\section{Numerical methods}

Our equilibria are constructed with the Cook-Shapiro-Teukolsky (CST) code
\cite{1992ApJ...398..203C}, which solves the Einstein equations for rotating
equilibria under the 
assumptions of stationarity and axisymmetry. The spacetime element 
(units of $G=c=1$) is in the form of
\begin{eqnarray}
ds^2 & = & -e^{\GG+\GR} dt^2 + e^{2\GA}(dr^2+r^2d\GU^2) \nonumber \\
     & + & e^{\GG-\GR}r^2\sin^2\GU (d\GP-\GO dt)^2 \ ,
\label{eq:ds2}
\end{eqnarray}
where $\GG,\GR,\GA,\GO$ are all functions of $r$ and $\GU$ only, while 
the stress-energy tensor is written as
\be
T_{\mu\nu} = (\GR_0 + \GR_i + P)u_\mu u_\nu + P g_{\mu\nu} \ ,
\label{eq:set}
\ee
where $\GR_0$ is the rest-mass density, $\GR_i$ the internal energy density, and $P$
the pressure at the rest frame of the fluid. Here, $u^\mu$ is the fluid four-velocity that
for a circular flow considered
here may be written as 
$u^\mu = u^t (t^\mu + \Omega \GP^\mu)$, where $t^\mu$ is the timelike Killing
vector that defines stationarity, and $\GP^\mu$ is the azimuthal spacelike
Killing vector that defines axisymmetry. 
The angular velocity $\Omega$ is constant for uniform
rotation but a function of $r$ and $\GU$ when differential rotation is considered.
The vanishing divergence of the stress-energy tensor, together with the assumptions
of stationarity and axisymmetry, lead to the Euler equation of hydrostatic equilibrium
\cite{1992ApJ...398..203C}. In the case of uniform rotation, the Euler equation 
can be directly integrated, while in the case of differential rotation, it can be 
integrated when the specific angular momentum $j=u^t u_\GP$ is a function of $\Omega$ itself 
\cite{1992ApJ...398..203C, Uryu:2016dqr}. 
In this work, we will consider either uniform rotation or differential rotation 
described by the Komatsu-Eriguchi-Hachisu law \cite{Komatsu:1989zz,1989MNRAS.239..153K}
$j(\Omega)= A^2 (\Omega_c - \Omega)$, where $\Omega_c$ is the angular velocity at the
center of the star, and $A$ is a parameter that controls the amount of differential
rotation. To probe for the existence of the ergosphere, we examine at every point 
the sign of the norm of the vector $t^\mu$, and in particular, we identify where the 
condition
\be
\mathbf{t}\cdot\mathbf{t} = g_{tt} = e^{\GG-\GR}(\GO^2 r^2 \sin^2\GU -e^{2\GR}) > 0 
\label{eq:gtt}
\ee
is satisfied.

\setlength{\tabcolsep}{5pt}                                                      
\begin{table}                                                             
\caption{The SLy EOS. In the first column are the dividing rest-mass 
densities in ${\rm g/cm^3}$, while in the second column are the polytropic indices.  } 
\label{tab:sly}                                                     
\begin{tabular}{cc}                                                    
\hline\hline                                     
$\GR_{0i}$              & $\Gamma_i$  \\ \hline\hline
-                       & $2.85100$   \\
$1.00000\times 10^{15}$ & $2.98800$   \\
$5.01187\times 10^{14}$ & $3.00500$   \\
$1.46220\times 10^{14}$ & $1.35692$ \\ \hline\hline
\end{tabular}                                                               
\end{table}                                                                   

The first EOS that we consider here uses the SLy EOS \cite{Douchin01} in the form of a 
piecewise representation $P=K_i\GR_0^{\Gamma_i}$ \cite{Read:2008iy}. The matching rest-mass 
densities $\GR_{0i}$ as well as the polytropic indices $\Gamma_i$ are shown in Table \ref{tab:sly}.
A polytropic constant is calculated from the reference values of pressure 
($2.42103\times 10^{34}\ {\rm dyn/cm^2}$)
and density ($5.01187\times 10^{14}\ {\rm g/cm^3}$), while the rest of the polytropic constants are
calculated from the equality of pressure at the dividing densities of Table \ref{tab:sly}.
The other three EOSs (SLycc1, SLycc2, and SLycc4) are based on the SLy one where we progressively 
substitute an inner core at matching densities $\GR_{0\rm nuc}, 2\GR_{0\rm nuc}$, and 
$4\GR_{0\rm nuc}$, with the maximally stiff EOS \cite{Haensel1989}
\be
P=\GS(\GE-\GE_s) + P_s \, .
\label{eq:eoscc}
\ee
Here, $\GS$ is a dimensionless parameter, $\GE=\GR_0+\GR_i$ is the total energy density,
and $P_s$ the pressure at $\GE_s$. The solutions presented in this work assume $\GS=1.0$,
i.e., a core at the causal limit, which represents the compressible EOS that yields 
configurations of maximal compactness
\cite{2016PhR...621..127L}. Equation (\ref{eq:eoscc}) relates the pressure to the total 
energy density when $\GR_0\geq\GR_{0s}=\GR_{0\rm nuc}, 2\GR_{0\rm nuc},4\GR_{0\rm nuc}$  
for the SLycc1, SLycc2, and SLycc4 EOSs respectively, while for $\GR_0\leq\GR_{0s}$ the SLy EOS 
is recovered. One can express the pressure in terms of the rest-mass density
in a polytropiclike form by integrating the first law of the thermodynamics
$d\GE/(\GE+P)=d\GR_0/\GR_0$. Using Eq. (\ref{eq:eoscc}), we get for $\GR_0\geq\GR_{0s}$,
\begin{eqnarray}
P   & = & \frac{1}{\GS+1}(\GS\GK \GR_0^{\GS+1} + P_s - \GS\GE_s) \, , \label{eq:precc}\\
\GE & = & \frac{1}{\GS+1}(\GK \GR_0^{\GS+1} + \GS\GE_s - P_s)    \, , \label{eq:rhocc}\\
h   & = & \GK \GR_0^\GS  \, ,  \label{eq:hcc} 
\end{eqnarray}
where the constant $\GK=h_s/\GR_{0s}^\GS$, and $h=(\GR_0 + \GR_i + P)/\GR_0$ is the specific enthalpy. 
The value $h_s$ can be evaluated from the polytrope outside the core.

\setlength{\tabcolsep}{5pt}                                                      
\begin{table}                                                             
\caption{The four EOSs employed here. The columns are the maximum spherical mass \protect \Mtov in units
of $M_\odot$, the maximum mass of the uniformly rotating models \protect \Mkep, and the corresponding 
rest-mass densities in ${\rm g/cm^3}$.  } 
\label{tab:eos}                                                     
\begin{tabular}{ccccc}                                                    
\hline\hline                                     
EOS     & \Mtov   & \Mkep   & \rhotov               & \rhokep                \\ \hline\hline
SLy     & $2.061$ & $2.488$ & $1.999\times 10^{15}$ & $1.771\times 10^{15}$  \\
SLycc1  & $4.067$ & $5.280$ & $5.979\times 10^{14}$ & $5.175\times 10^{14}$  \\
SLycc2  & $2.917$ & $3.656$ & $1.139\times 10^{15}$ & $1.022\times 10^{15}$  \\
SLycc4  & $2.222$ & $2.681$ & $1.852\times 10^{15}$ & $1.721\times 10^{15}$  \\ \hline\hline
\end{tabular}                                                               
\end{table}                                                                   

\section{Results}

For the four EOSs described above the maximum spherical mass \Mtov, the maximum mass at
the mass-shedding limit under uniform rotation \Mkep, as well as their corresponding rest-mass densities are 
shown in Table \ref{tab:eos}. We note here that the SLy EOS has a speed of sound 
$c_s=\sqrt{dP/d\GE}$ larger than the speed of light when $\GR_0>= 1.999\times 10^{15}\ {\rm g/cm^3}$
which is identical to the density at the maximum mass. For the other three EOSs,  
since $4\GR_{0\rm nuc}$ is less than this value, we always have $c_s\leq c$.

\begin{figure*}
\begin{center}
\includegraphics[width=0.68\columnwidth]{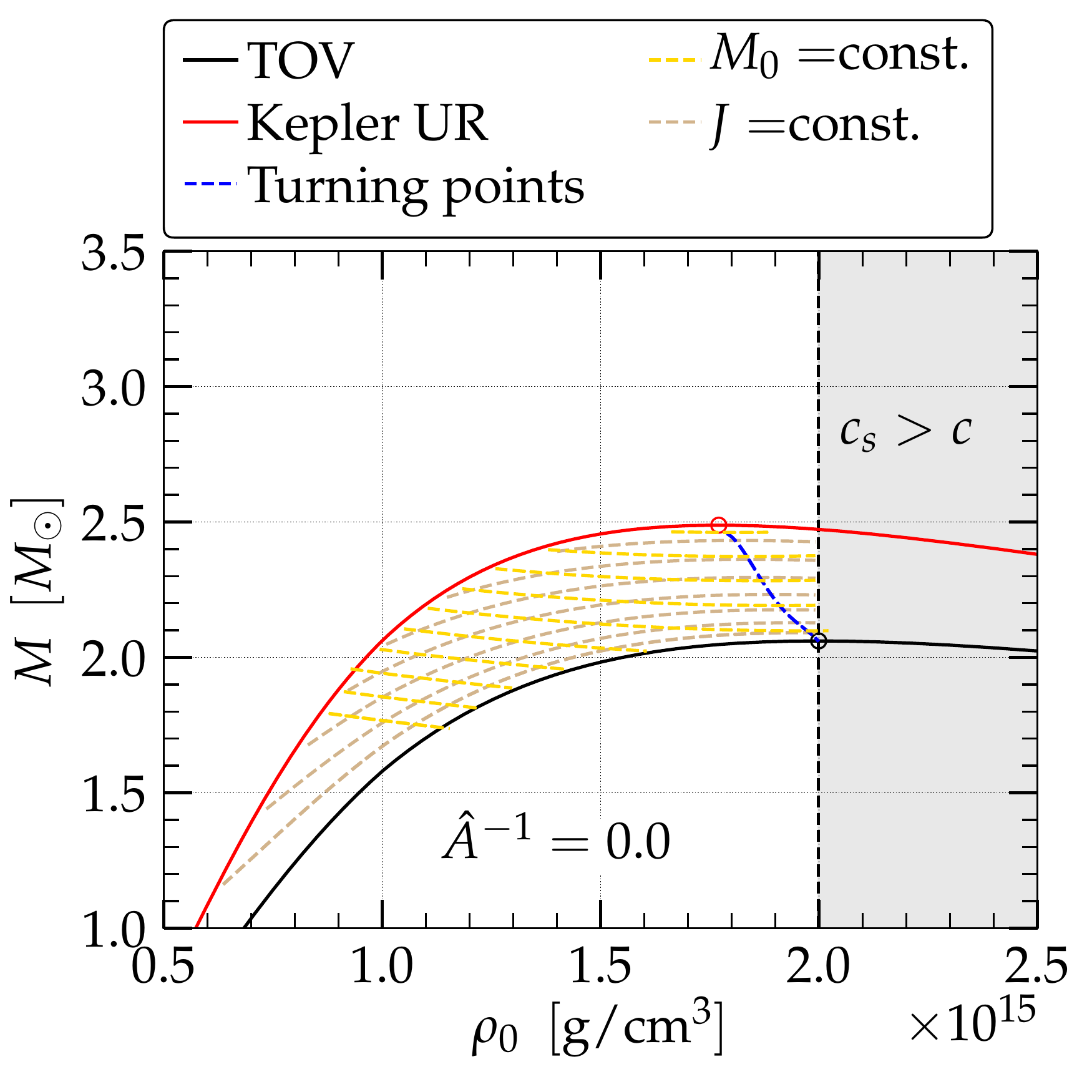}
\includegraphics[width=0.68\columnwidth]{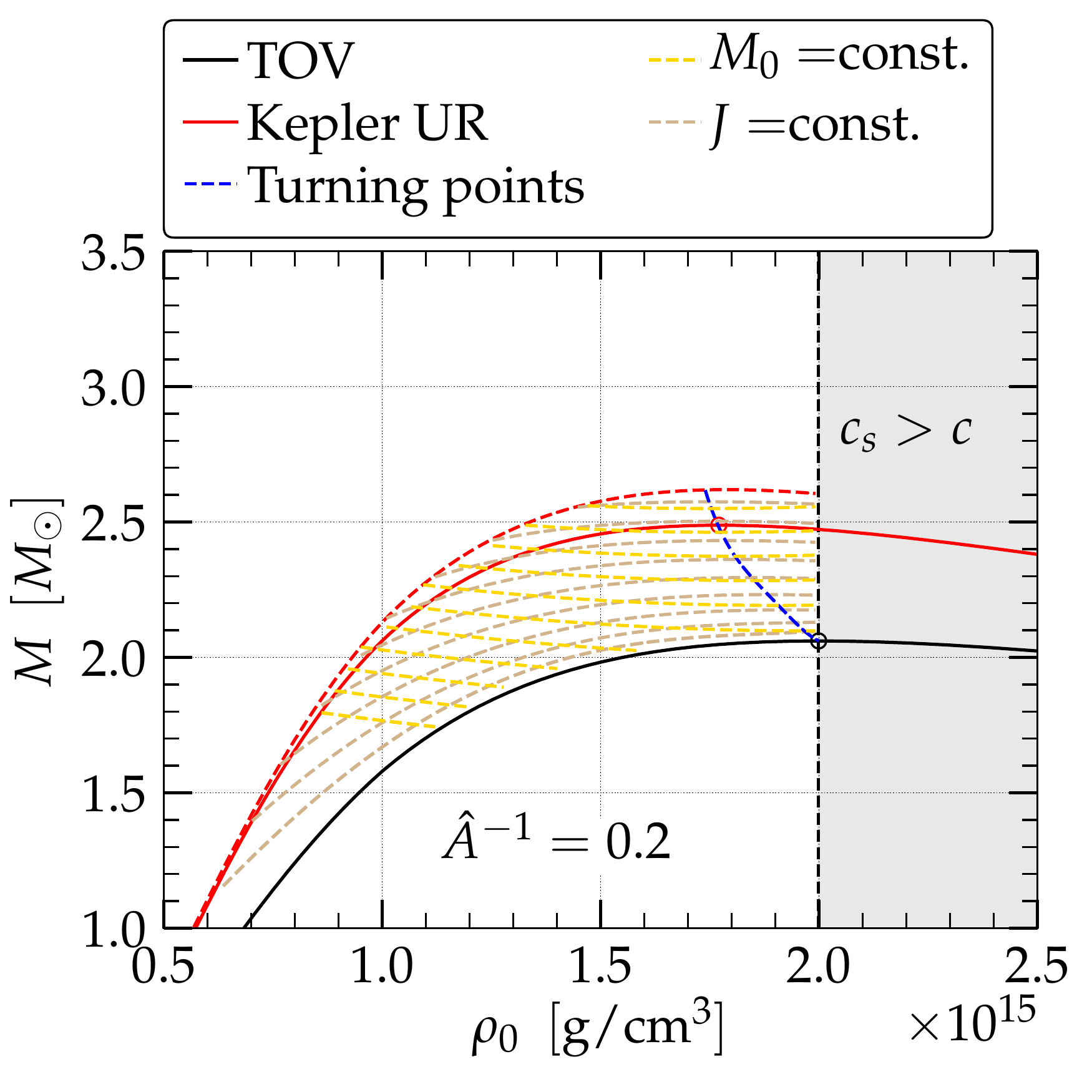}
\includegraphics[width=0.68\columnwidth]{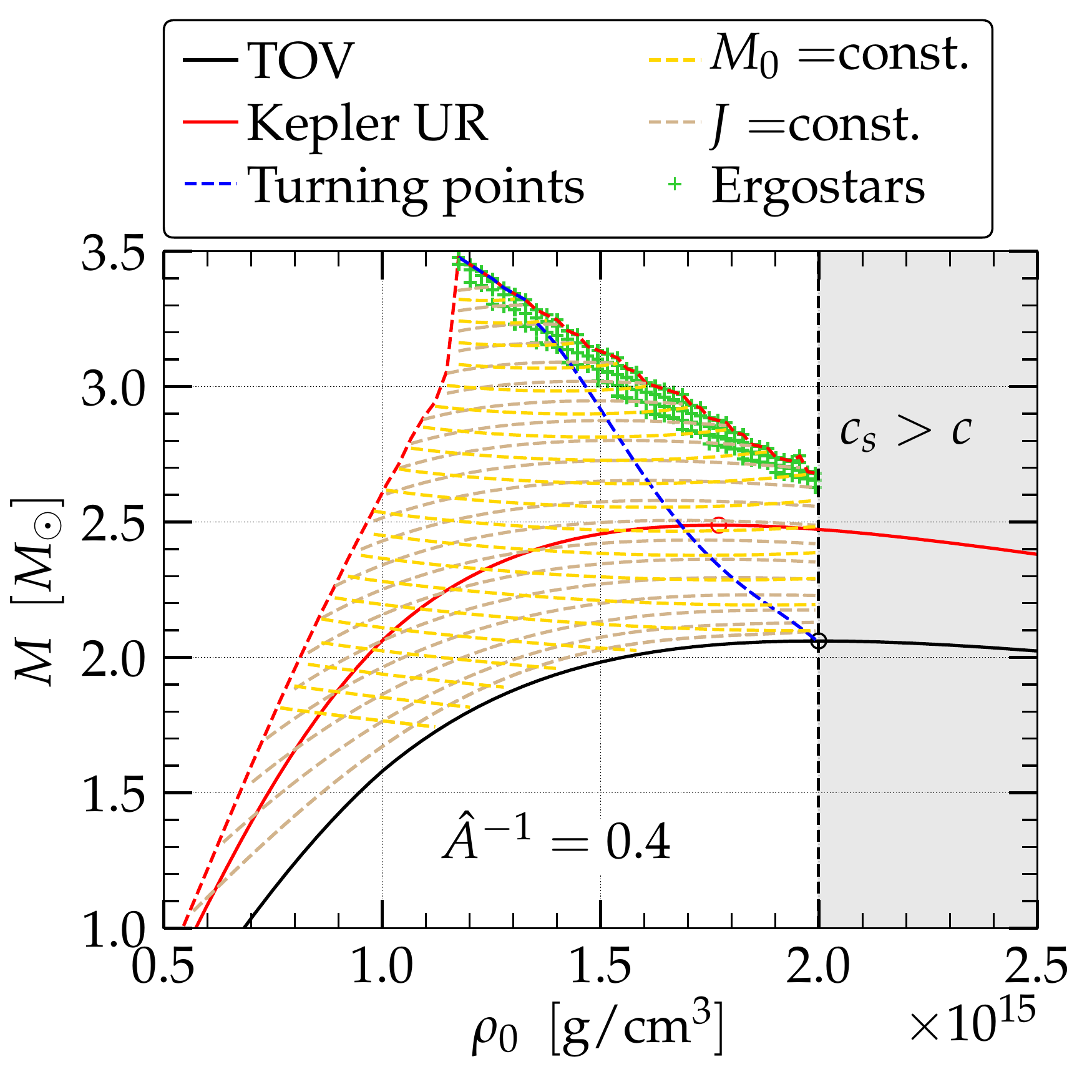}
\includegraphics[width=0.68\columnwidth]{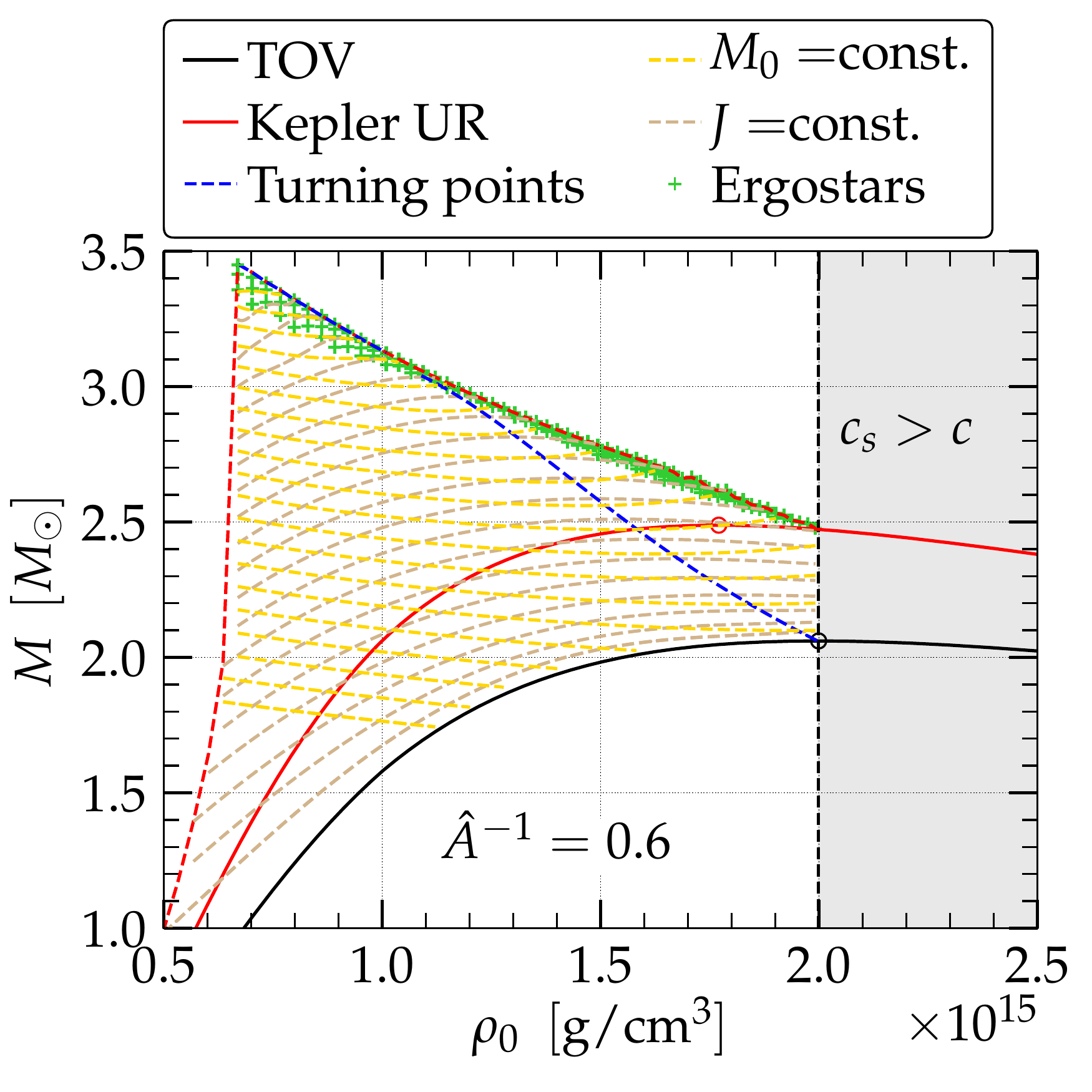}
\includegraphics[width=0.68\columnwidth]{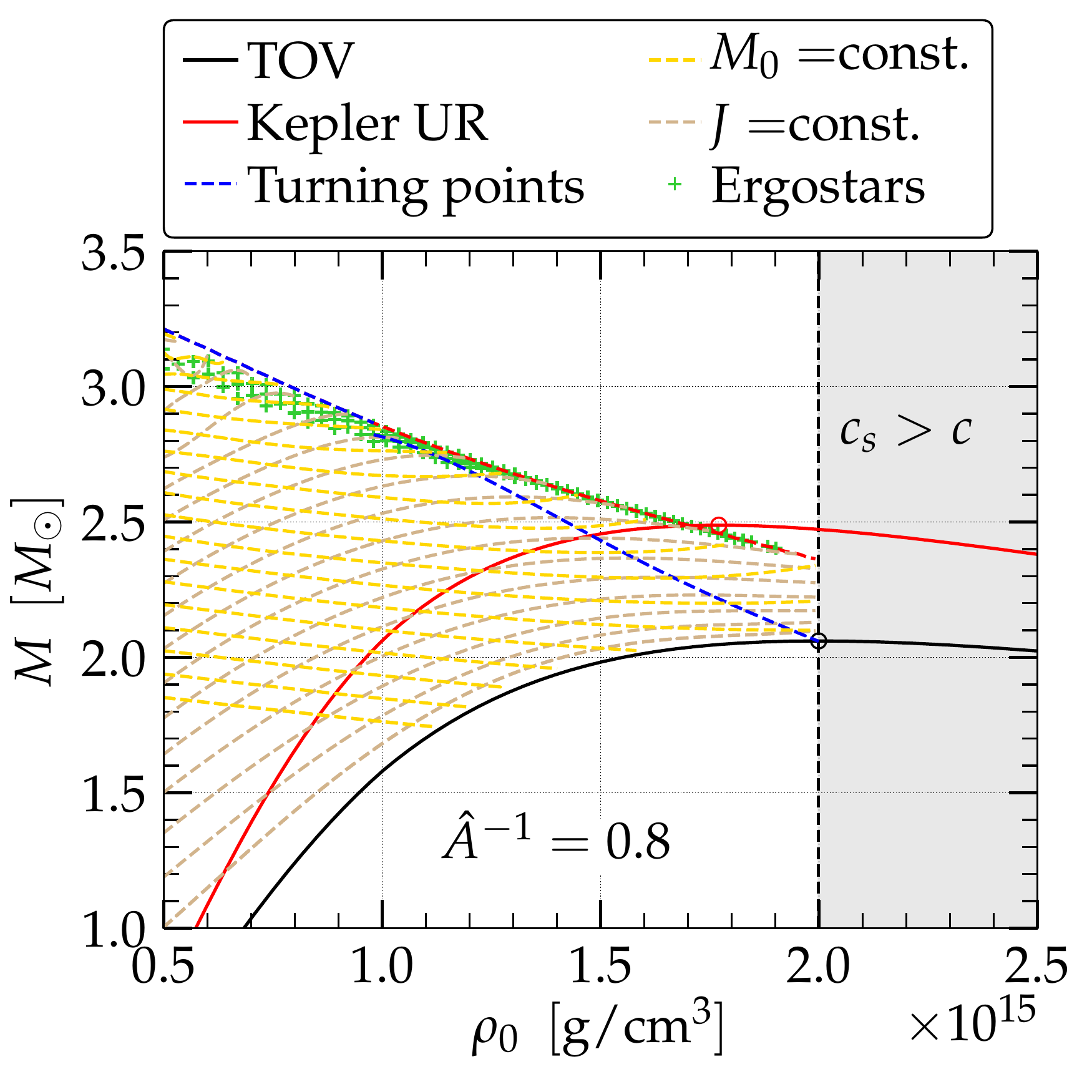}
\includegraphics[width=0.68\columnwidth]{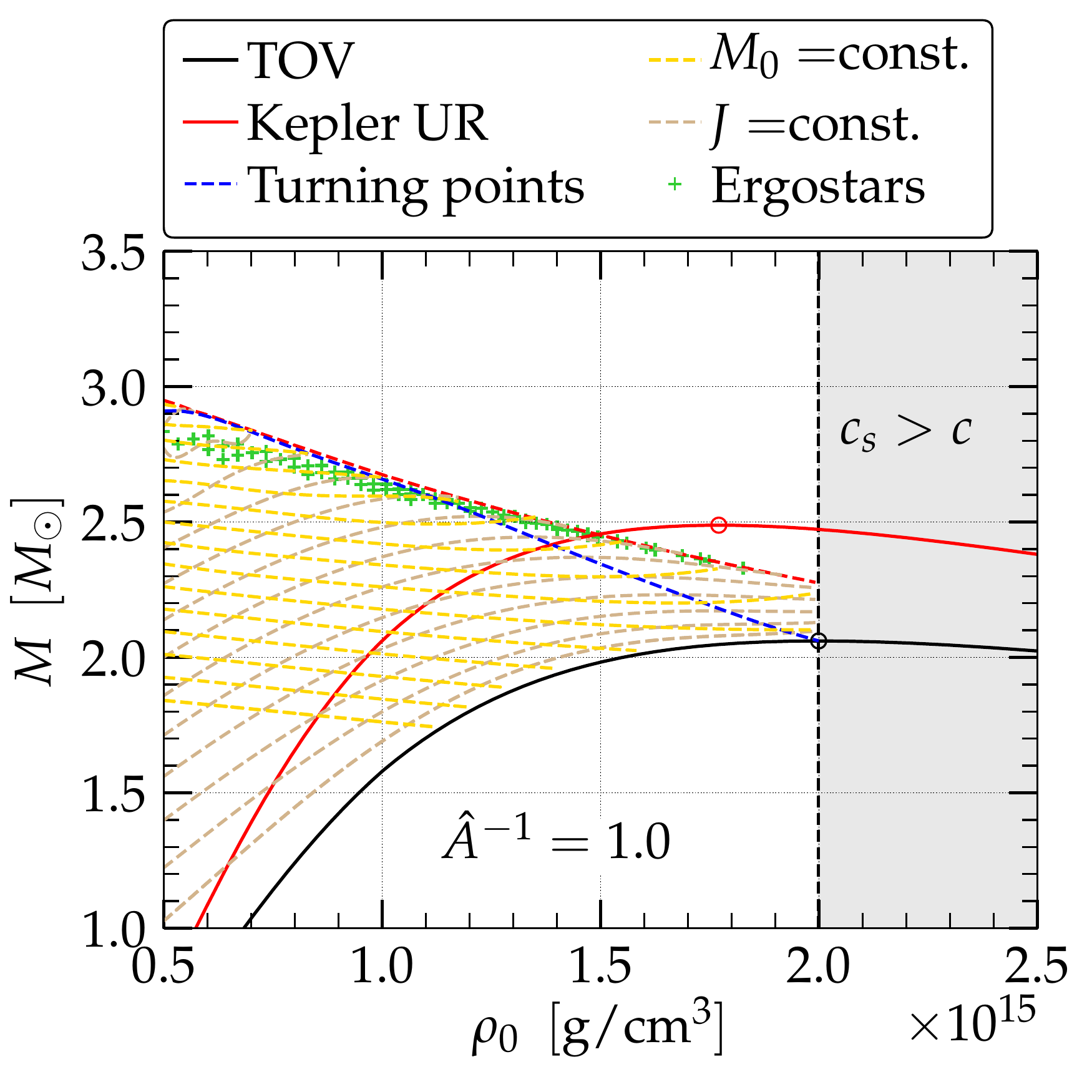}
\includegraphics[width=0.68\columnwidth]{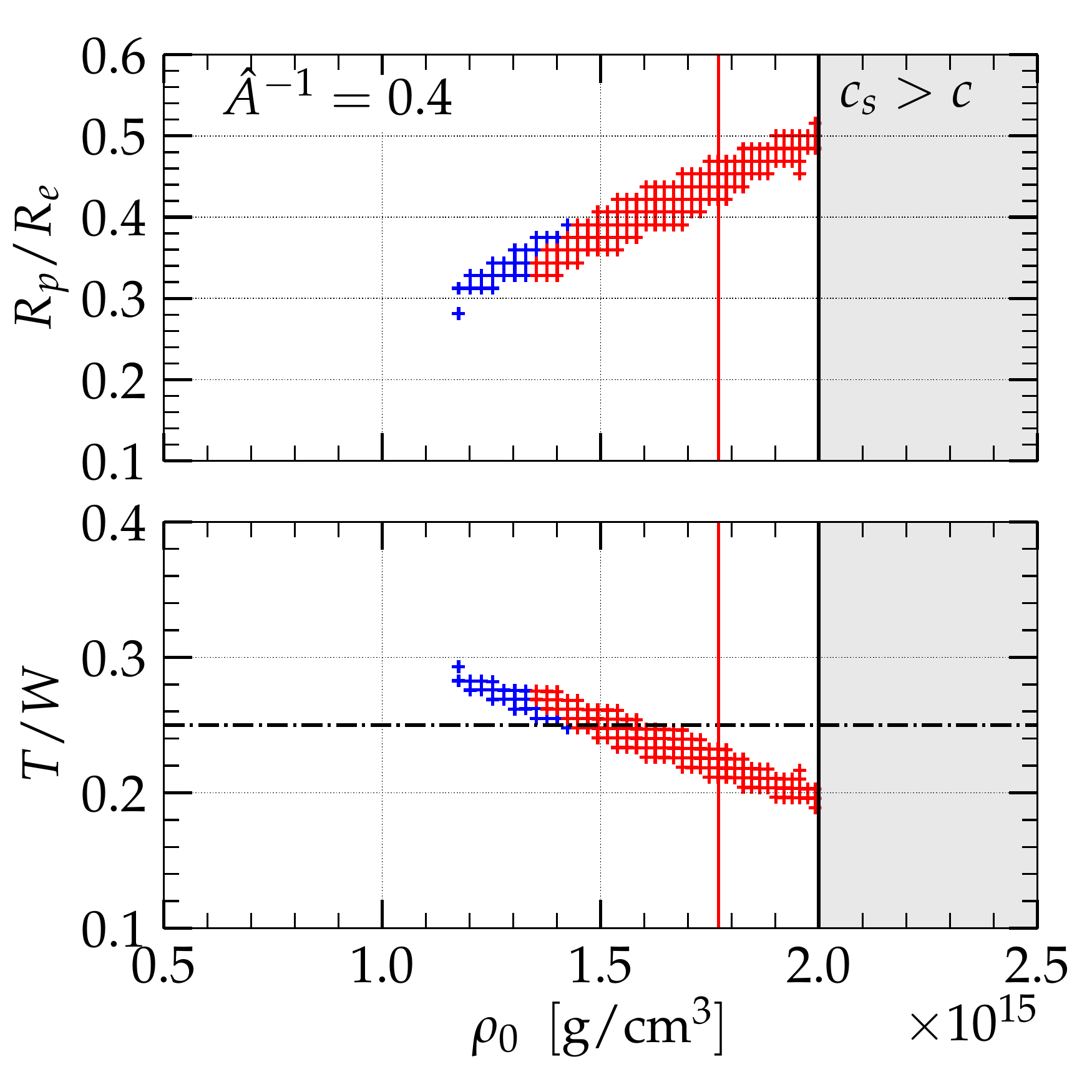}
\includegraphics[width=0.68\columnwidth]{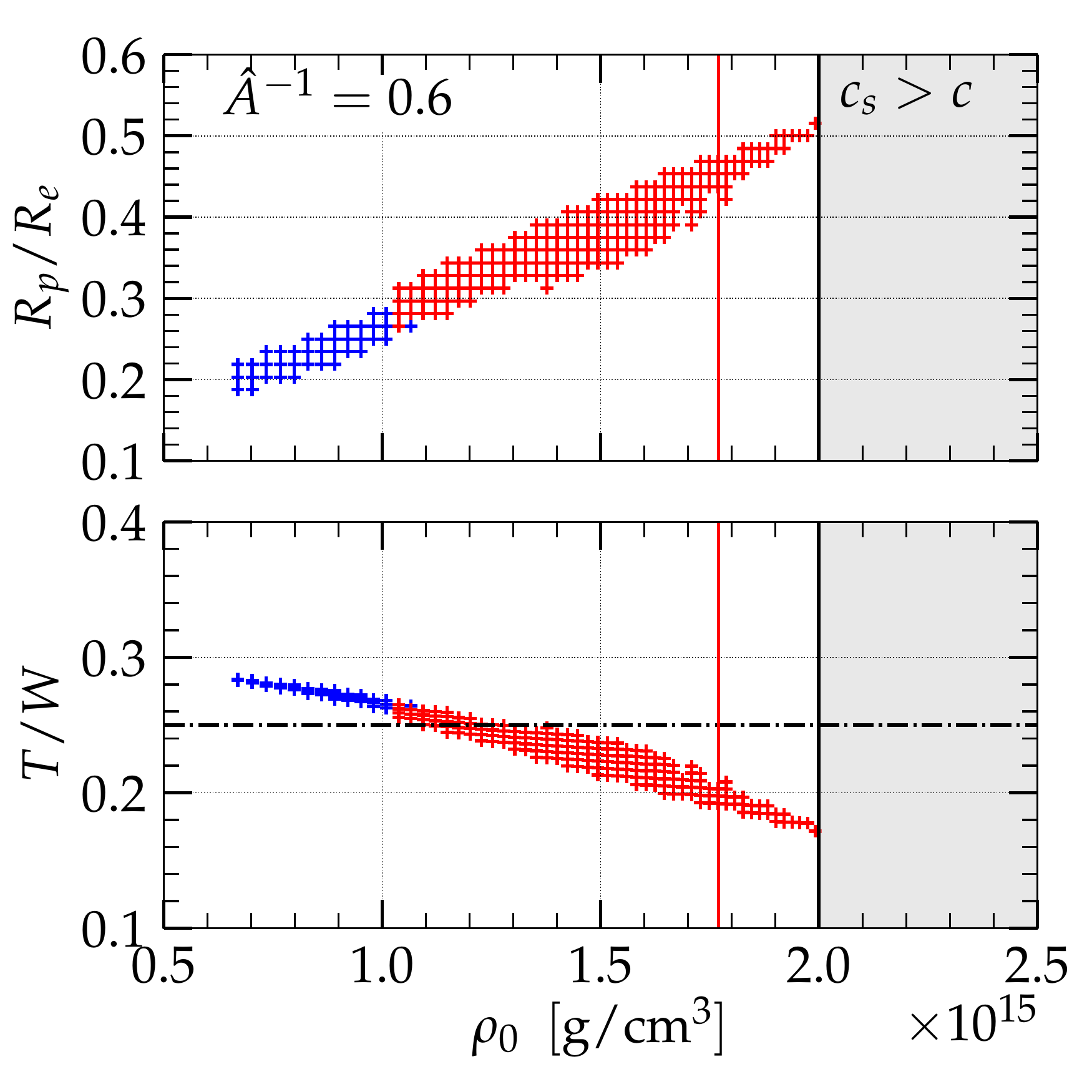}
\includegraphics[width=0.68\columnwidth]{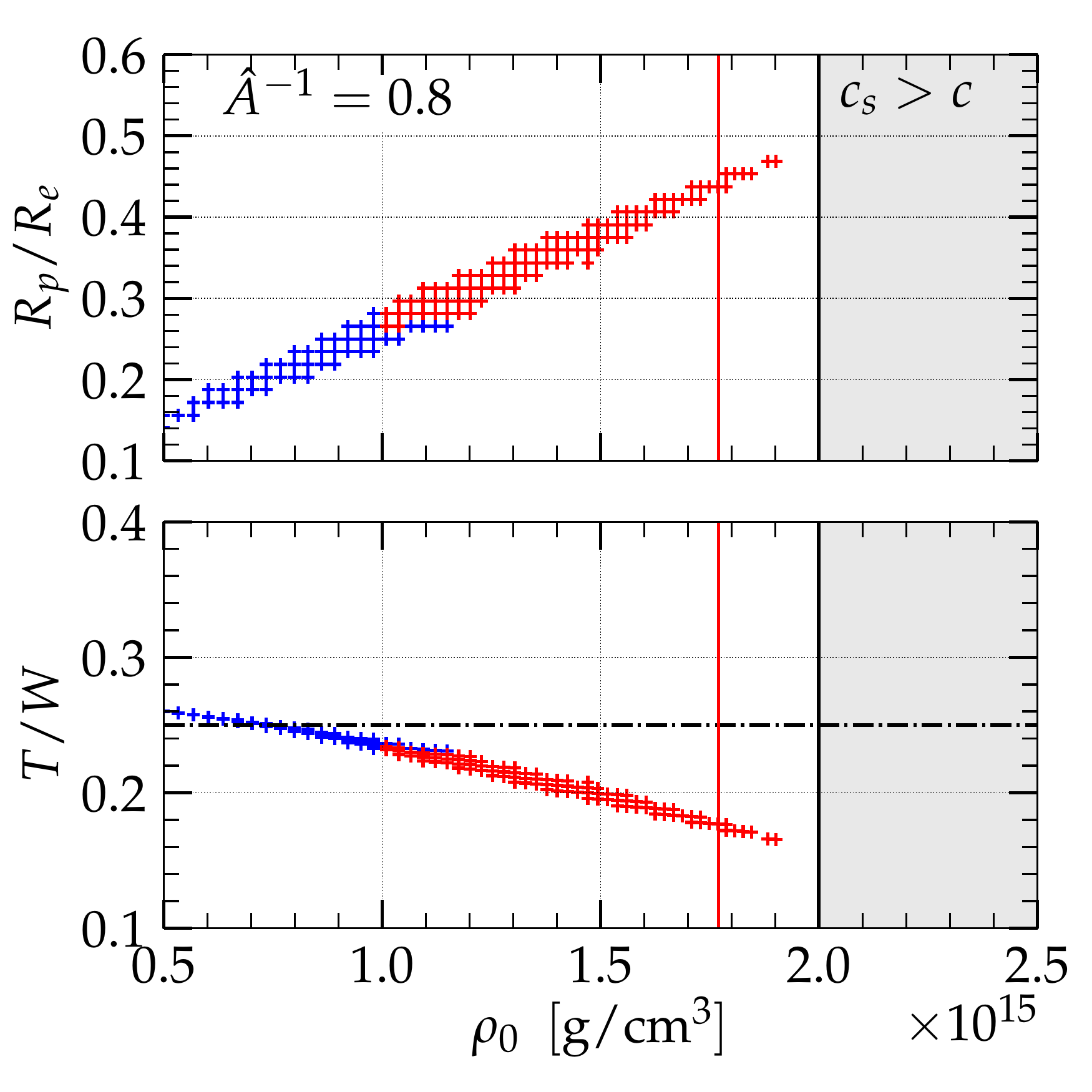}
\caption{SLy EOS. Top and middle row depict the location of the ergostars (green cross)
in a mass vs central rest-mass density diagrams for different degrees of differential rotation $\hat{A}$. 
The panel with $\hat{A}^{-1}=0$ (top left) corresponds to uniform rotation. Bottom row 
shows the deformation ($R_p/R_e$) and $T/W$ for three differential rotation cases.
Blue crosses correspond to ergostars on the left of the turning point line, while red crosses
correspond to the ones on the right.}
\label{fig:SLy}
\end{center}
\end{figure*}

\begin{figure*}
\begin{center}
\includegraphics[width=0.68\columnwidth]{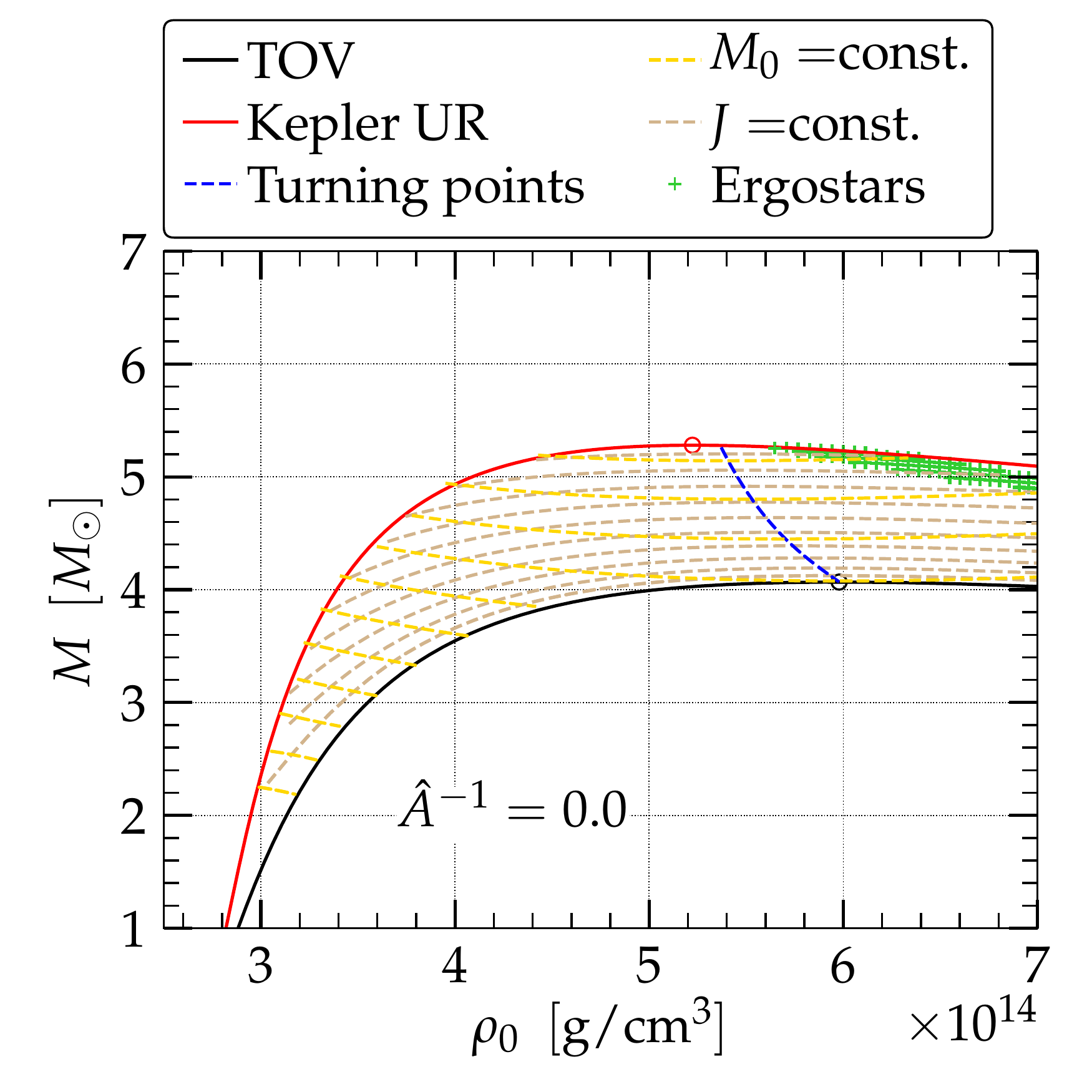}
\includegraphics[width=0.68\columnwidth]{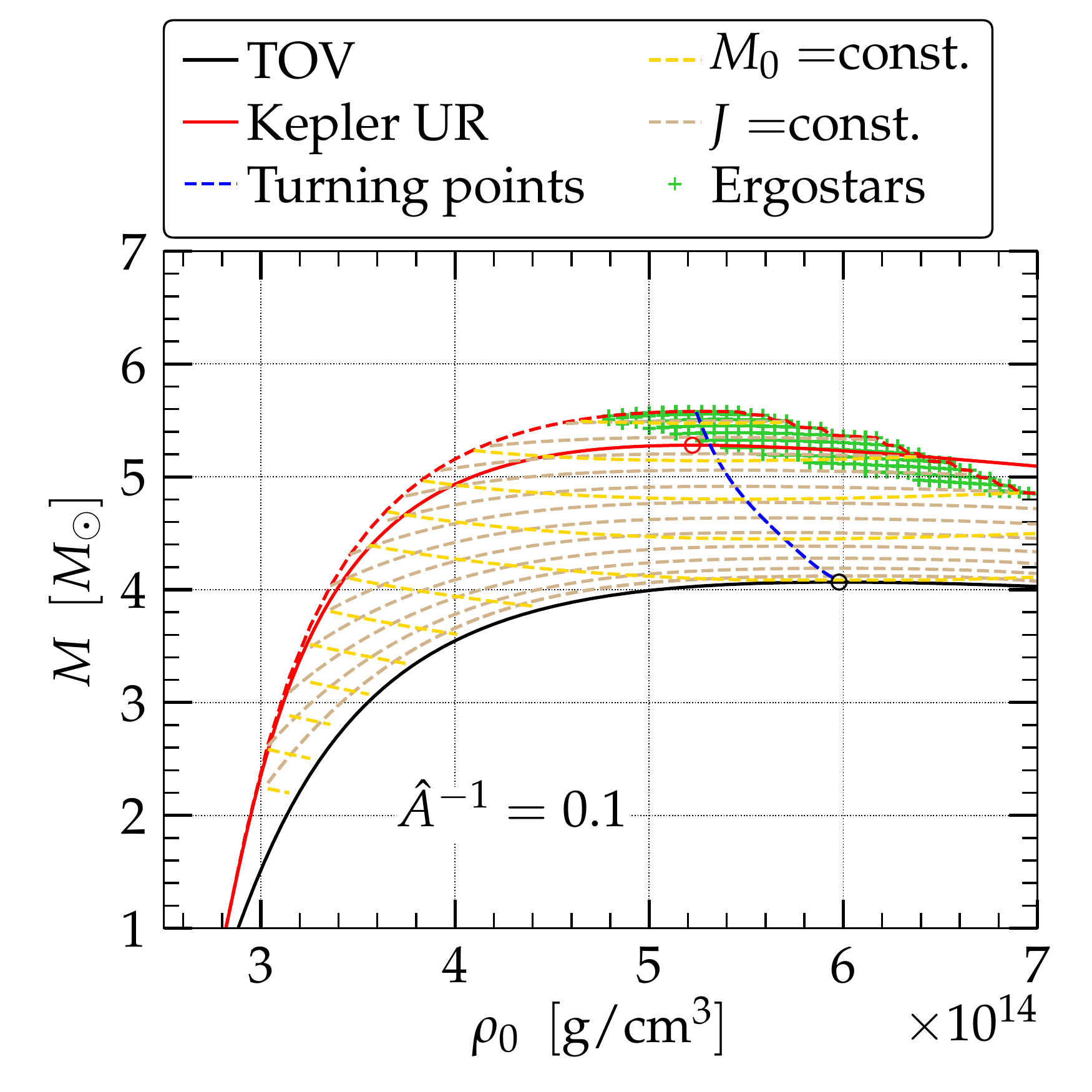}
\includegraphics[width=0.68\columnwidth]{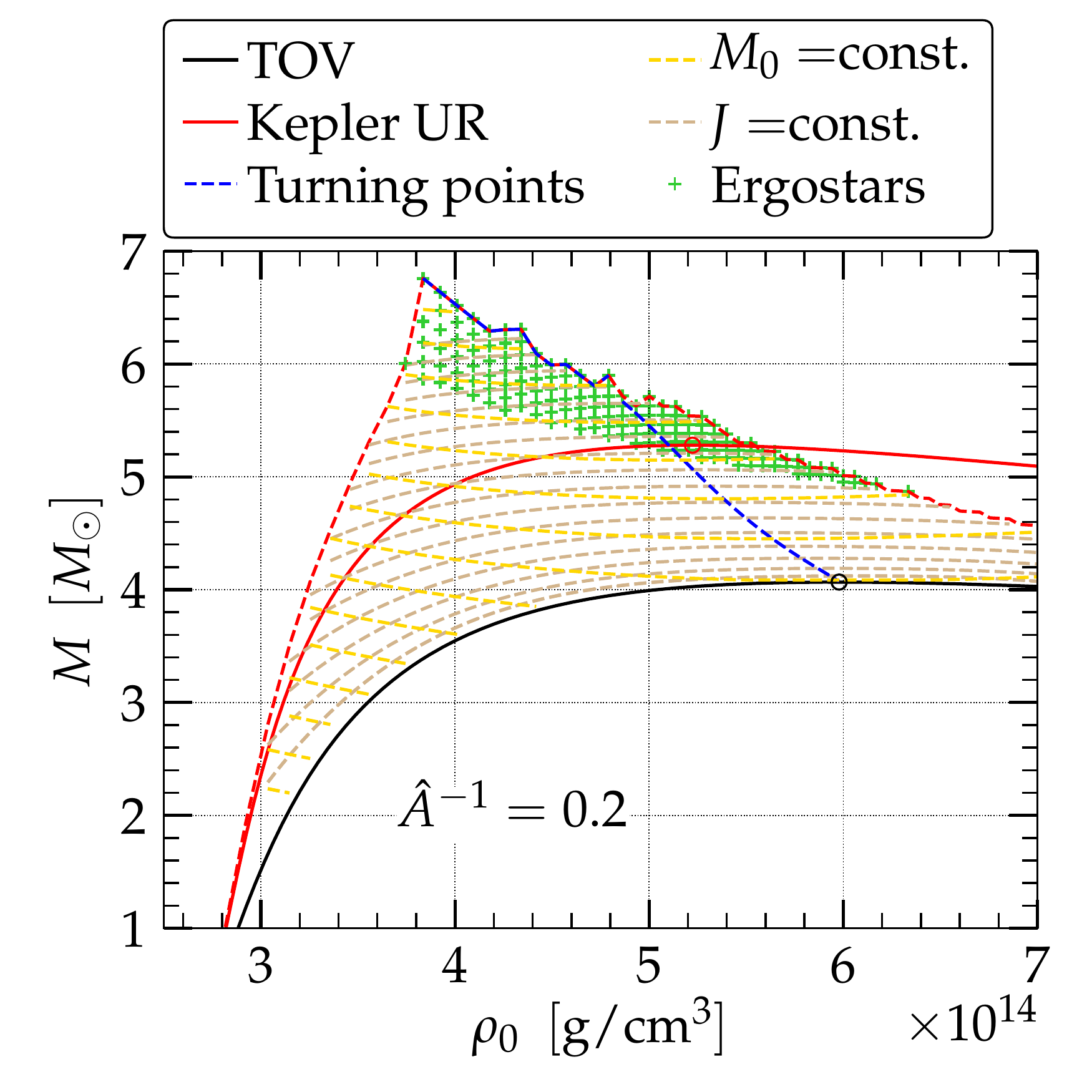}
\includegraphics[width=0.68\columnwidth]{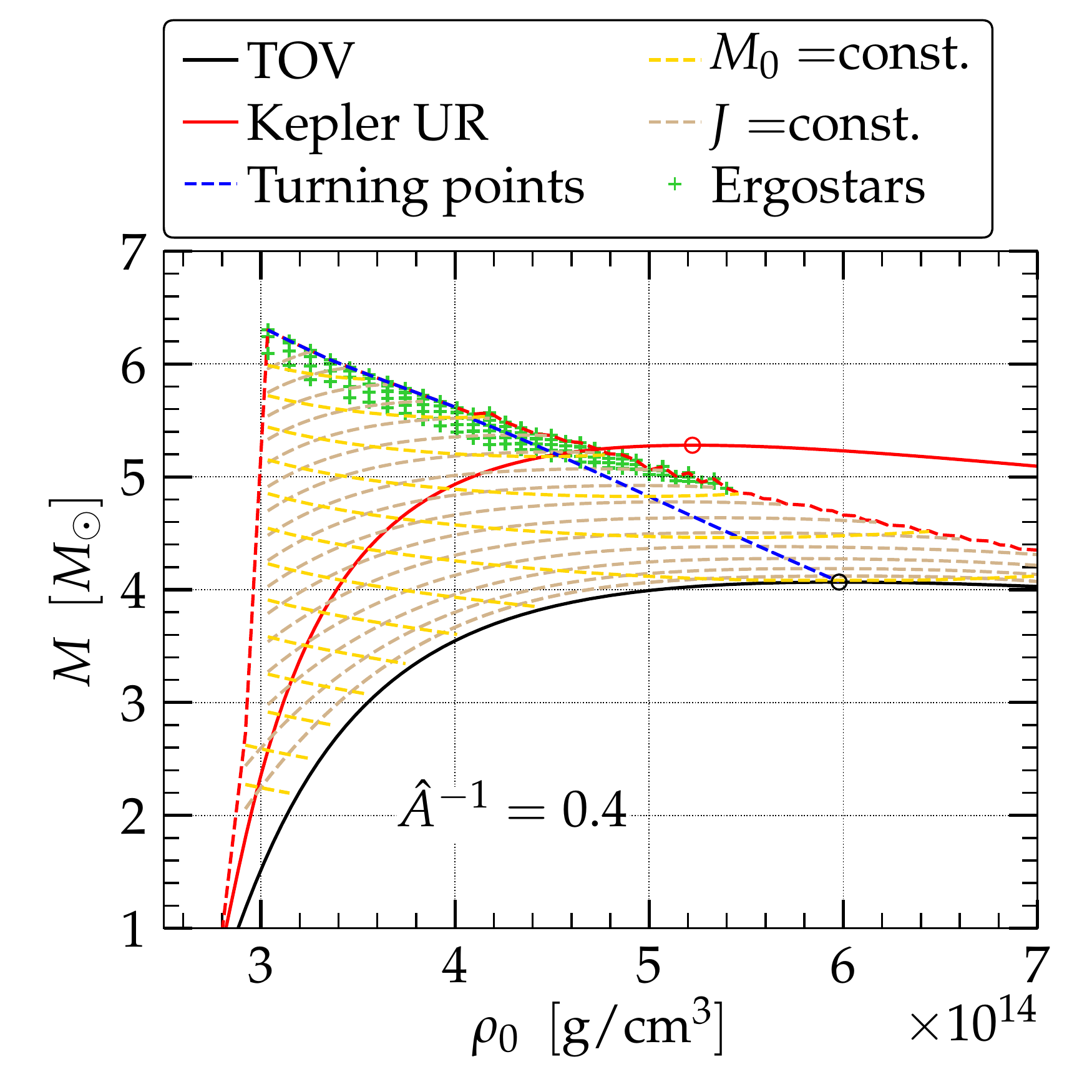}
\includegraphics[width=0.68\columnwidth]{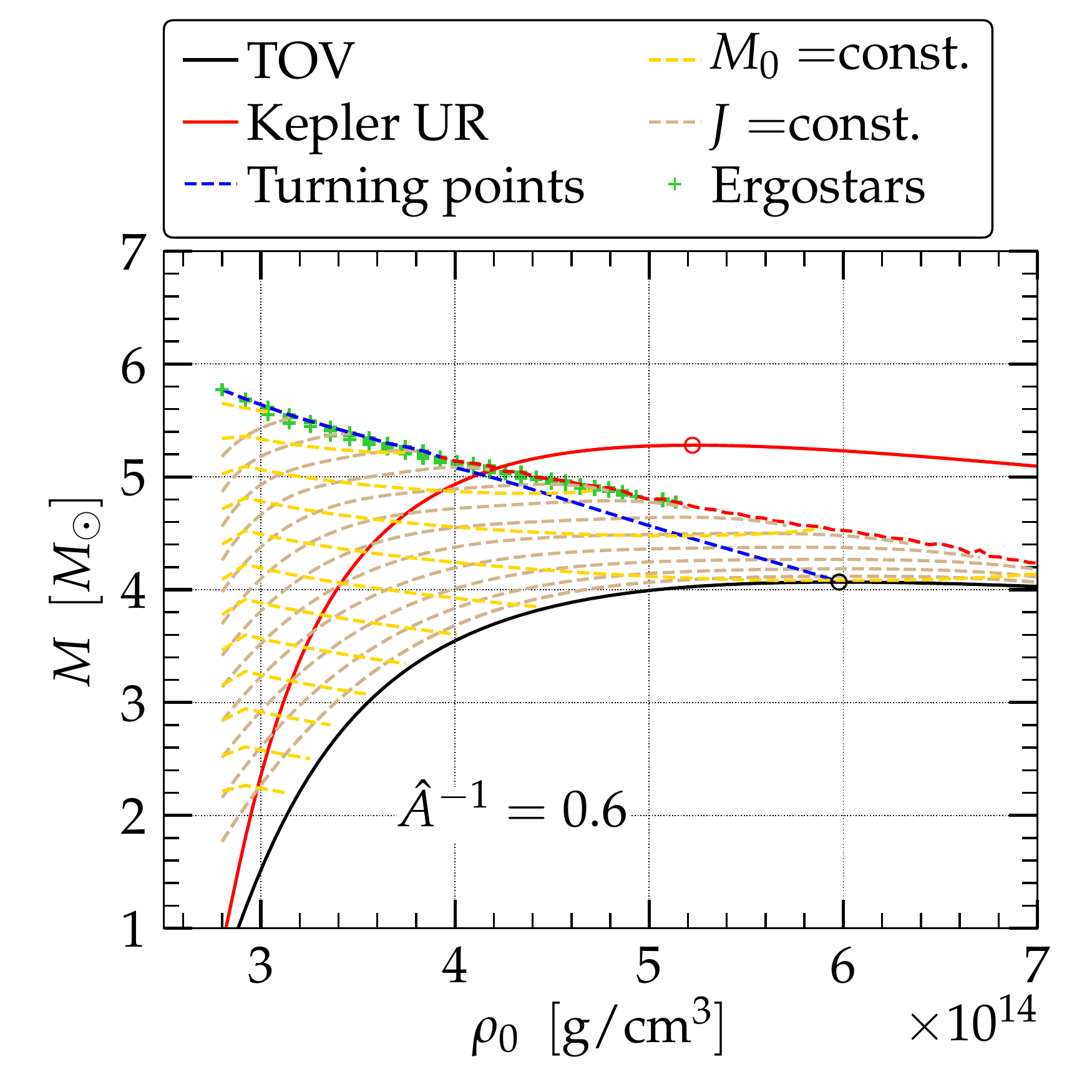}
\includegraphics[width=0.68\columnwidth]{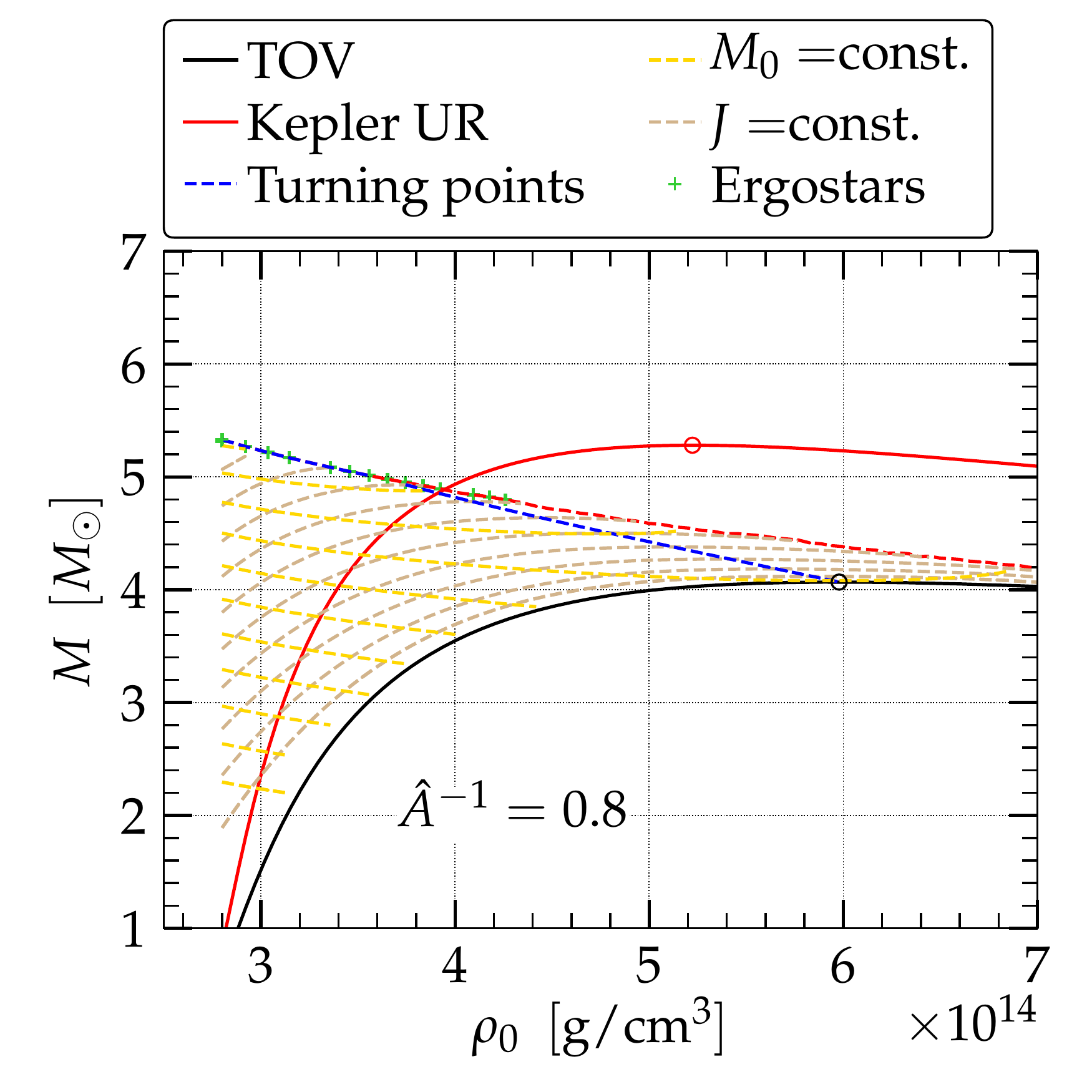}
\includegraphics[width=0.68\columnwidth]{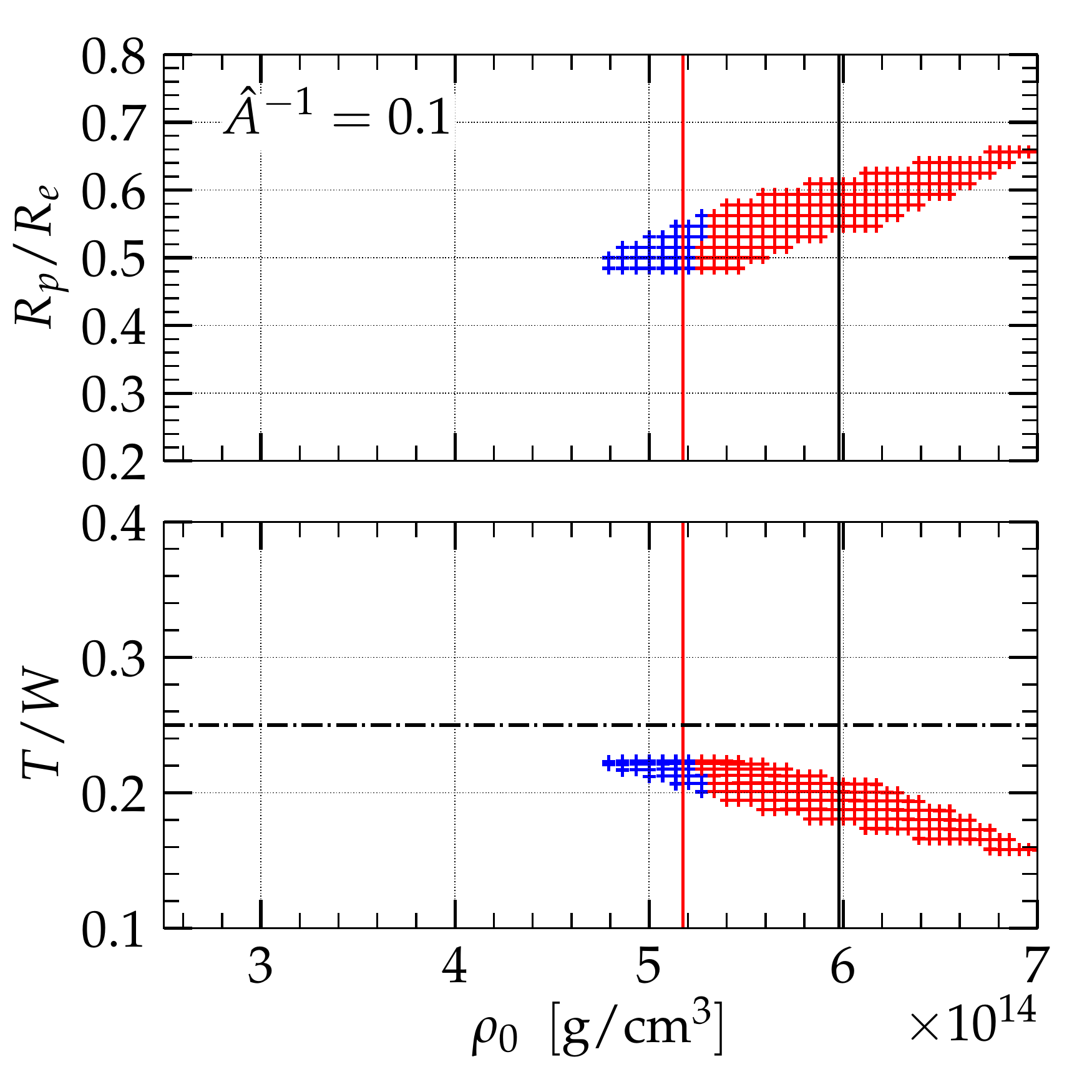}
\includegraphics[width=0.68\columnwidth]{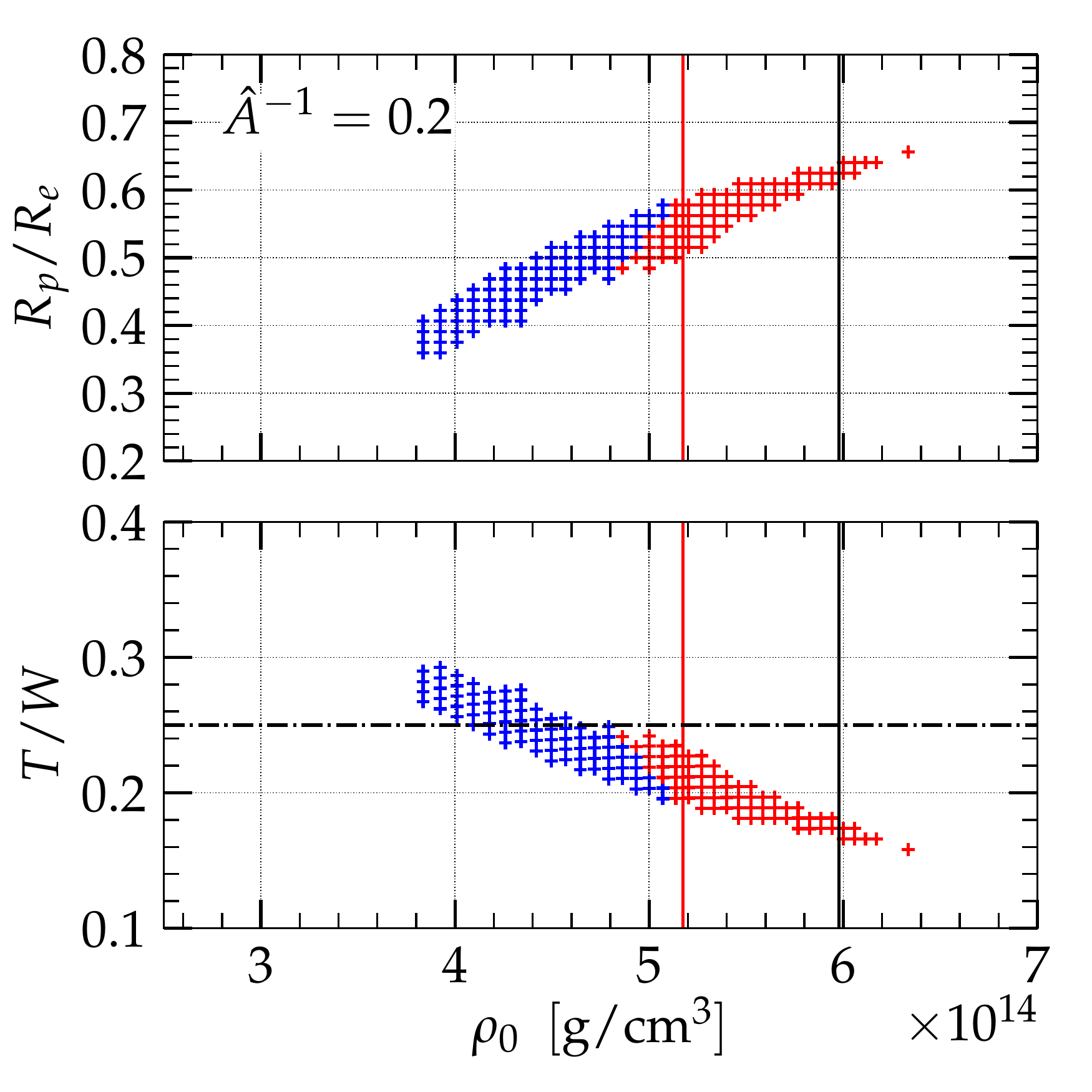}
\includegraphics[width=0.68\columnwidth]{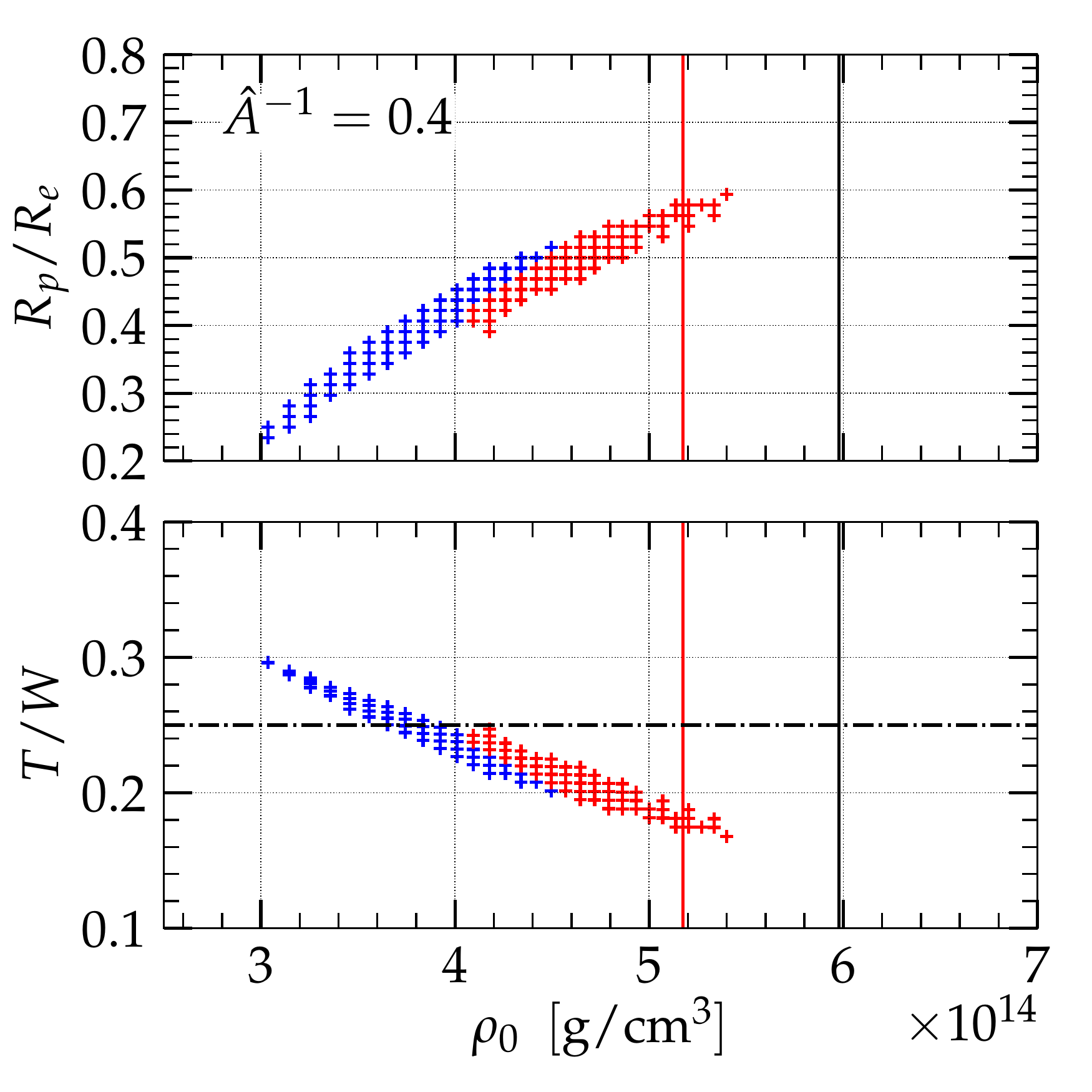}
\caption{Same as Fig. \ref{fig:SLy} but for the SLycc1 EOS.}
\label{fig:SLycc1}
\end{center}
\end{figure*}

Figure \ref{fig:SLy} is devoted to the SLy EOS. The top and middle rows depict the position of
ergostars (green crosses) in a mass vs central rest-mass 
density diagram.\footnote{Here we use the notation of the CST code
\cite{1992ApJ...398..203C} where $\hat{A}=A/R_e$, $R_e$ being the equatorial radius.}
Every panel corresponds to a different degree of differential rotation
starting from uniform rotation in the top left panel where $\hat{A}^{-1}=0$ and progressing
to a higher degree of differential rotation in the right middle panel where $\hat{A}^{-1}=1$. In 
each plot, we show the spherical solutions (TOV black curve), the mass-shedding limit of uniformly 
rotating stars (red curve), sequences of constant rest mass $M_0$ (orange dashed curves), sequences 
of constant angular momentum $J$ (brown dashed curves), and the curve that joins the maximum mass 
points (turning points) 
on every $J={\rm const}$ sequence (blue dashed curve). In a typical calculation, for every 
$\hat{A}$ (i.e., for every panel)
we divide a range of densities starting from a low density up to the limiting point
$2.0\times 10^{15}\ {\rm g/cm^3}$ into $60$ intervals, and using the CST code we compute $60$ 
constant rest-mass density sequences from the spherical limit (black curve) all the way up into
more massive models that have small ratios $R_p/R_e$ until the code fails to converge. 
Here, $R_p$ is the polar radius.
The last points, i.e., the points with the smaller value of $R_p/R_e$, on every sequence are 
connected with a dashed red curve in the panels of the top and middle row in Fig. \ref{fig:SLy}. 
As we can see, there are no ergostars for uniformly rotating models or small differential rotation
$\hat{A}^{-1}=0.2$ for the SLy EOS. On the other hand, the largest number of ergostars appear
when $\hat{A}^{-1}$ is approximately in the range $0.4-0.6$, while for larger degrees of 
differential rotation, they tend to diminish again. 

One important line in these plots is the blue dashed line which separates the secularly unstable/stable
models against axisymmetric perturbations. For the uniformly rotating case, it is denoted as the turning 
point line due to ``turning point theorem'' of Friedman, \textit{et al.} \cite{1988ApJ...325..722F}: 
Along a sequence of uniformly rotating stars with fixed angular momentum and increasing central density, 
the configuration of maximum mass marks the onset of \textit{secular} instability. The turning point line
is also commonly taken to be the criterion for distinguishing \textit{dynamical} stability. Although the 
analysis of Takami \textit{et al.} \cite{2011MNRAS.416L...1T} implies that the loci of secular, dynamical, and
turning point lines is more subtle, they clearly are close to each other. For differential rotation,
there is no analogous theorem, but there is significant evidence that again the locus of dynamical 
stability is very close to the turning point 
on $J=$const curves \cite{Kaplan:2013wra,Bauswein:2017aur,Weih:2017mcw}.
According to \cite{2011MNRAS.416L...1T,Weih:2017mcw}, the dynamical instability typically sets in at 
central densities slightly below the one that corresponds to the turning point. In particular,
as one moves along a $J=$const sequence toward increasing densities, one encounters the
secular instability first, then the dynamical, and finally the turning point.
Given that all three points are very close together and 
given the lack of any general theorem, we will
assume here that the turning points mark the beginning of the dynamically unstable region, although
the reader should be aware of the differences discussed above.
We also note here that in the cases with differential rotation, the CST code can go to large deformations
(i.e., small ratios of $R_p/R_e$) that correspond to toroidal configurations. On the other, hand 
in some cases, especially for large masses and smaller densities, we were not able to find a turning point. 
Typically, for those cases a fixed angular momentum sequence is a monotonically increasing function of mass 
as one moves to larger densities. For our present purposes, we tacitly assume that the last points in 
those sequences signify the dynamical instability 
limit, although in reality that limit should be on the left at higher densities.\footnote{It is possible
that more fine-tuned codes like 
\cite{Stergioulas:1994ea,Ansorg_2009} can go beyond our calculated models and refine the position 
of the turning point in the very high mass differentially rotating regime.}

This implies that \textit{all ergostars on and to the right of the blue dashed lines are dynamically unstable}. 
For the SLy EOS, Fig. \ref{fig:SLy}, this criterion rules out most of the ergostars, at least for a
mild degree of differential rotation ($\hat{A}^{-1}=0.4,0.6$). For larger differential rotation,
the ergostars tend to accumulate close to the turning point line (or more precisely, close to the last model
we were able to calculate), and given the discussion above,
the dynamical stability of these models is questionable, as a full evolution
will be needed for a diagnosis. We also note here that as differential rotation becomes larger, the
turning point line becomes straighter and rotates counterclockwise with respect to the maximum
spherical point. This also implies that 
\textit{all models on and to the right of the uniformly rotating
turning point line are dynamically unstable irrespective of the degree of differential rotation}.
In addition, this is true for supramassive as well as hypermassive stars.

To get a better understanding of the qualitative features of the 
SLy ergostars, we plot in the bottom row of Fig. \ref{fig:SLy} the deformation parameter $R_p/R_e$ 
as well as the rotational kinetic over the gravitational potential energy $T/W$ for three representative 
cases of differential rotation. The vertical black line corresponds to the density of the maximum spherical
mass, while the red one corresponds to the density of the maximum uniformly rotating mass. We find that all models
with $\hat{A}^{-1}=0.4,0.6,0.8$ are toroidal (i.e., the maximum density is not at the center), 
and the larger the differential rotation, the more toroidal
shapes we were able to compute. Note that according to recent studies \cite{Espino:2019xcl},
extreme toroidal configurations are dynamically unstable. In the $T/W$ panels, we draw
with a horizontal dashed-dot line the $T/W=0.25$ benchmark, which in many cases provides a crude criterion
for the onset of dynamical instability to nonaxisymmetric (bar) modes \cite{Shibata_2000, Baumgarte:1999cq}. 
Blue crosses correspond to ergostars on the left of the turning point line, while red crosses correspond 
to ergostars on the right of the turning point line.

Figure \ref{fig:SLycc1} is similar to Fig. \ref{fig:SLy}, but it corresponds to the SLycc1 EOS. The effect of 
the large causal core is immediately seen even for the uniformly rotating models: Supramassive ergostars
now appear, but they all lie in the dynamical unstable part of the parameter space. We also note here that
the maximum mass of the spherical solutions, as well as the maximum mass at the mass-shedding limit, increase
considerably from the SLy EOS (by factors of $1.97$ and $2.12$, respectively). This has already been seen
with the ALF2cc EOS employed in \cite{Tsokaros:2019mlz, Tsokaros:2019lnx}. Given the fact that the SLycc1
and ALF2cc EOSs only differ in the crust 
(i.e., for $\GR_0\leq \GR_{0\rm nuc}$), it is not surprising that the differences in the TOV and Kepler lines are 
minute. In addition, comparing SLycc1 vs SLy, 
we see that the densities where the maximum mass for the spherical and mass-shedding
sequence reduce significantly (by factors of $3.34$ and $3.42$, respectively; see Table \ref{tab:eos}).
Although this may seem contradictory, it is related to the fact that the density profile along an axis
is quite different from models using a typical EOS (like SLy or ALF2) without a large causal core. 
Instead of a parabolic type, 
the profile with SLycc1 starts from a smaller central density, diminishes somewhat all the way to the surface 
of the star, where it abruptly reduces to zero (see Fig. 1 in \cite{Tsokaros:2019lnx}). In this respect,
stars with the SLycc1 or ALF2cc EOS resemble quark stars that have a finite surface density. 

\begin{figure*}
\begin{center}
\includegraphics[width=0.68\columnwidth]{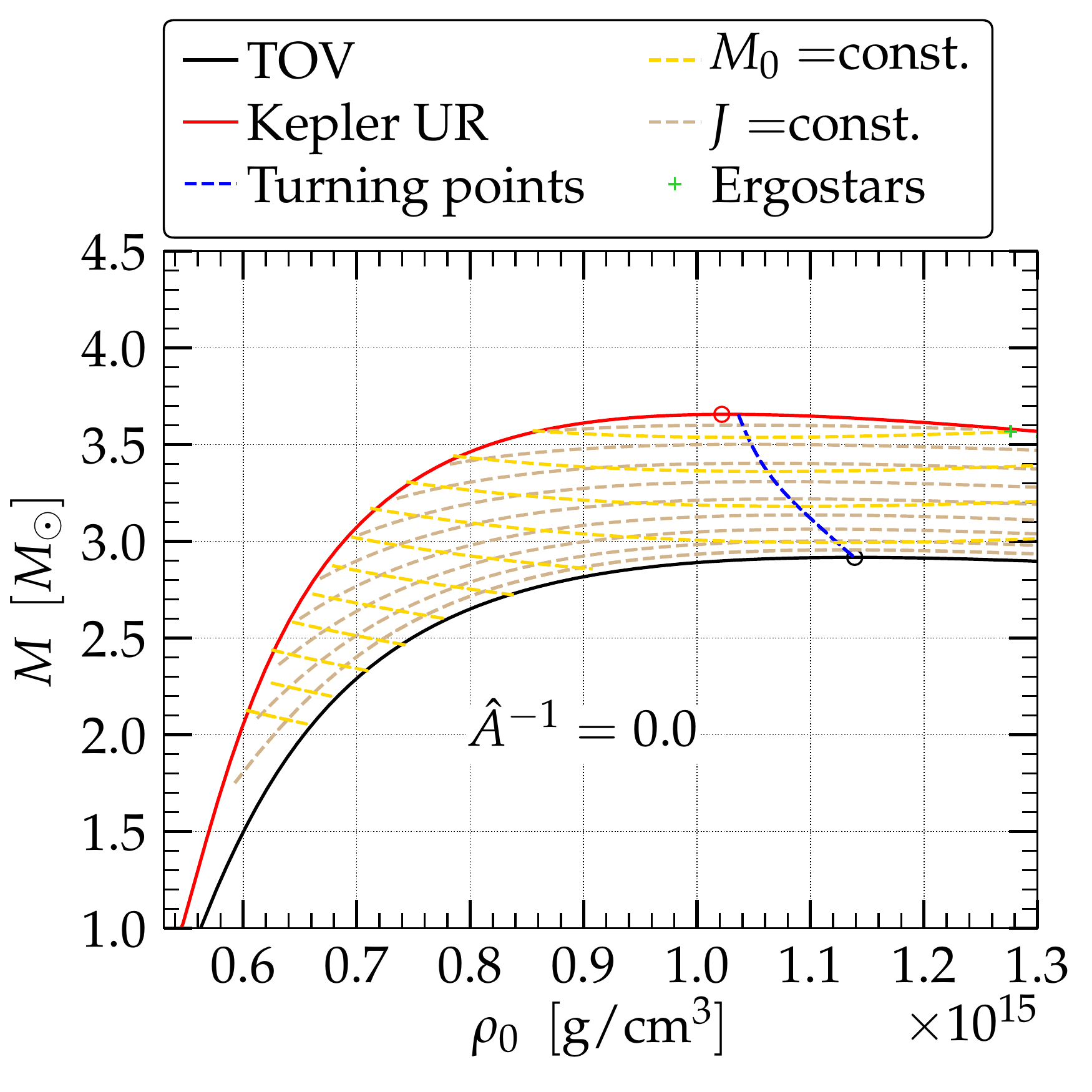}
\includegraphics[width=0.68\columnwidth]{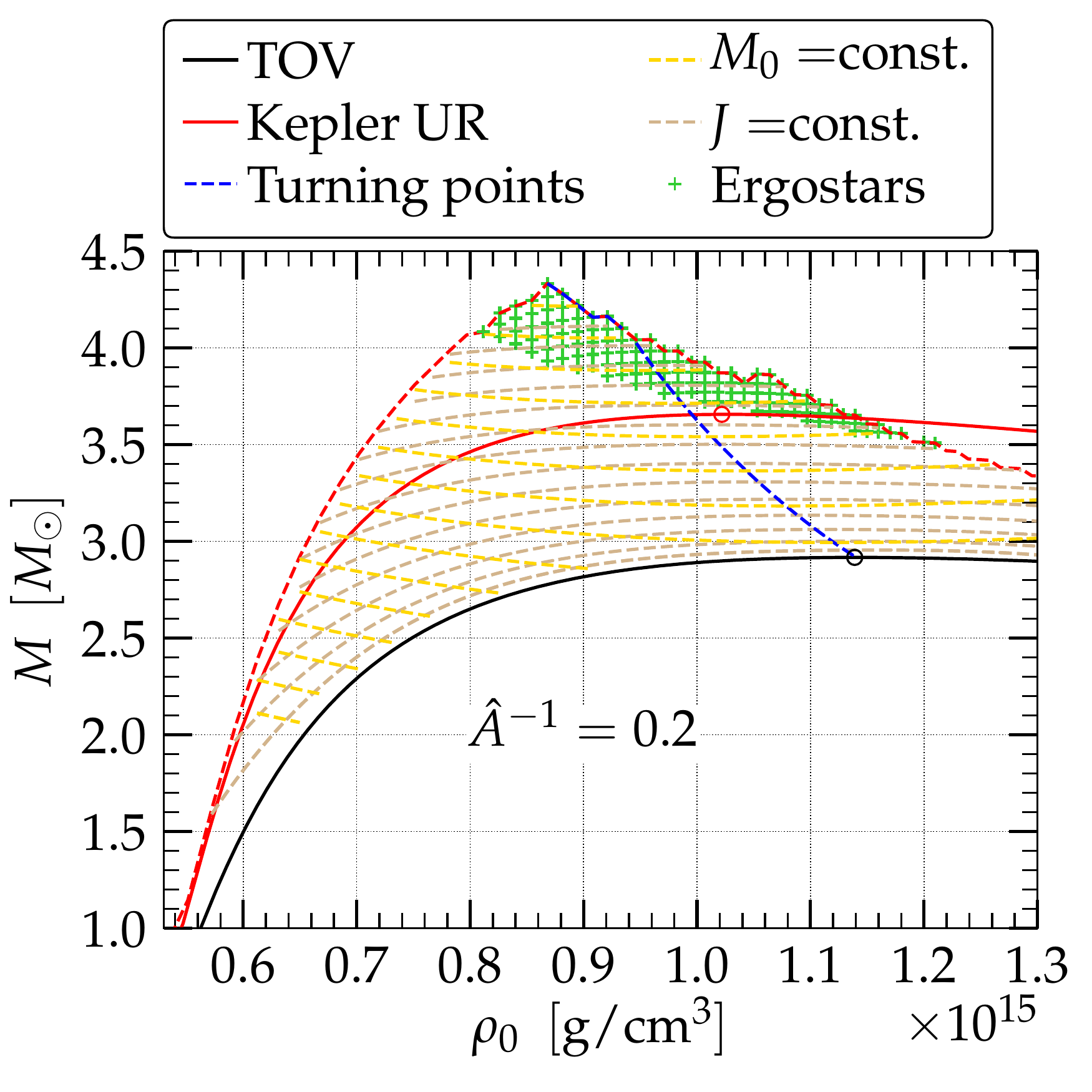}
\includegraphics[width=0.68\columnwidth]{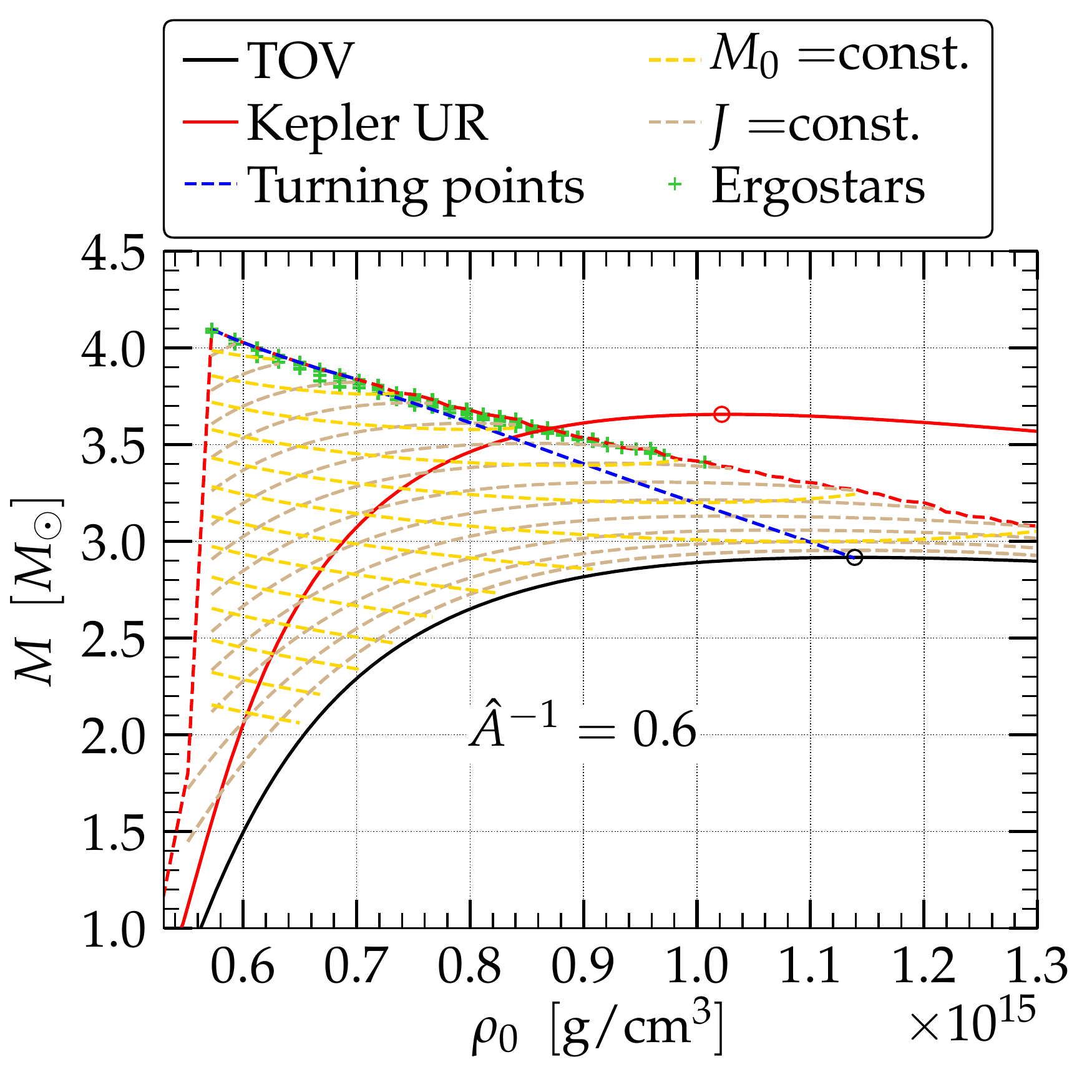}
\includegraphics[width=0.68\columnwidth]{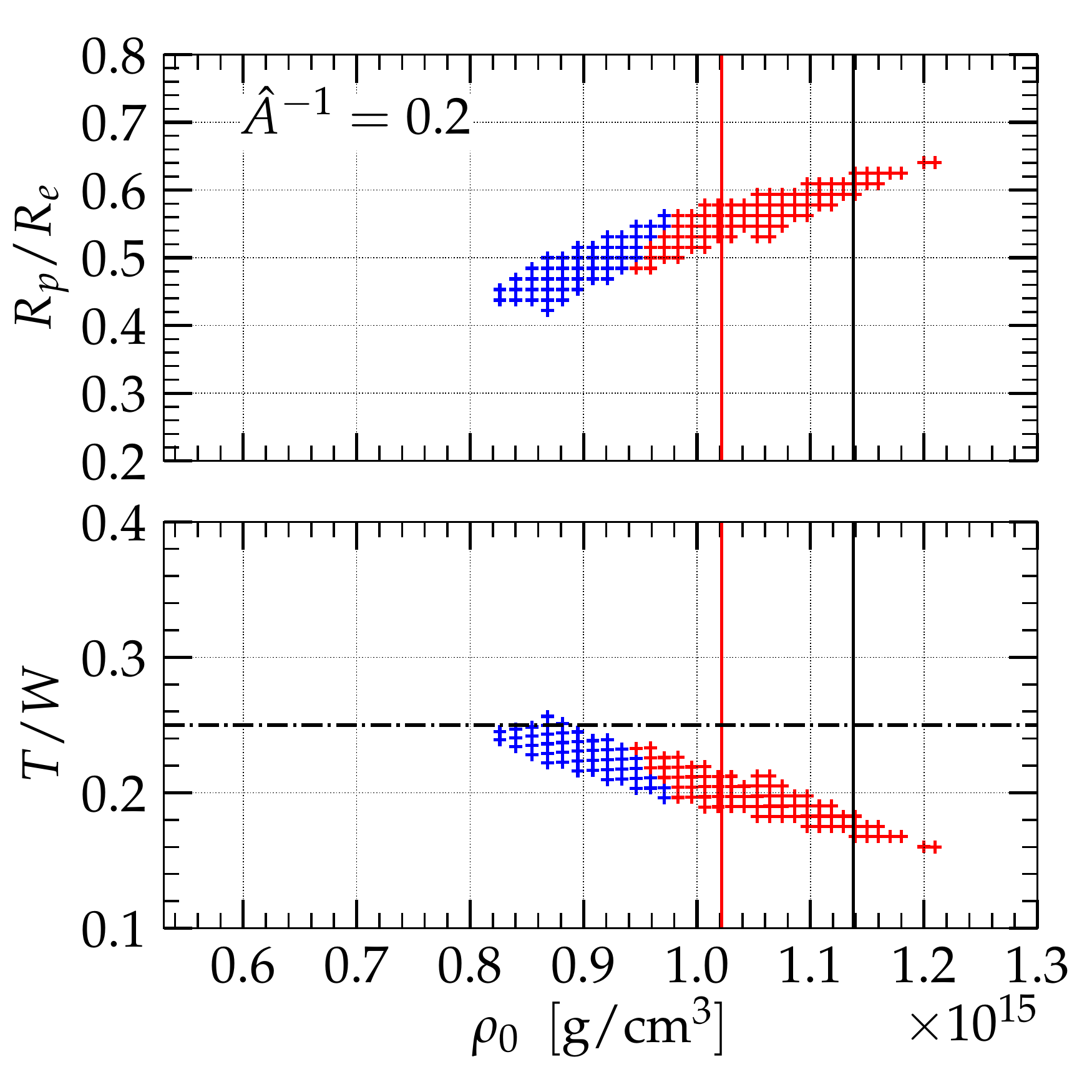}
\includegraphics[width=0.68\columnwidth]{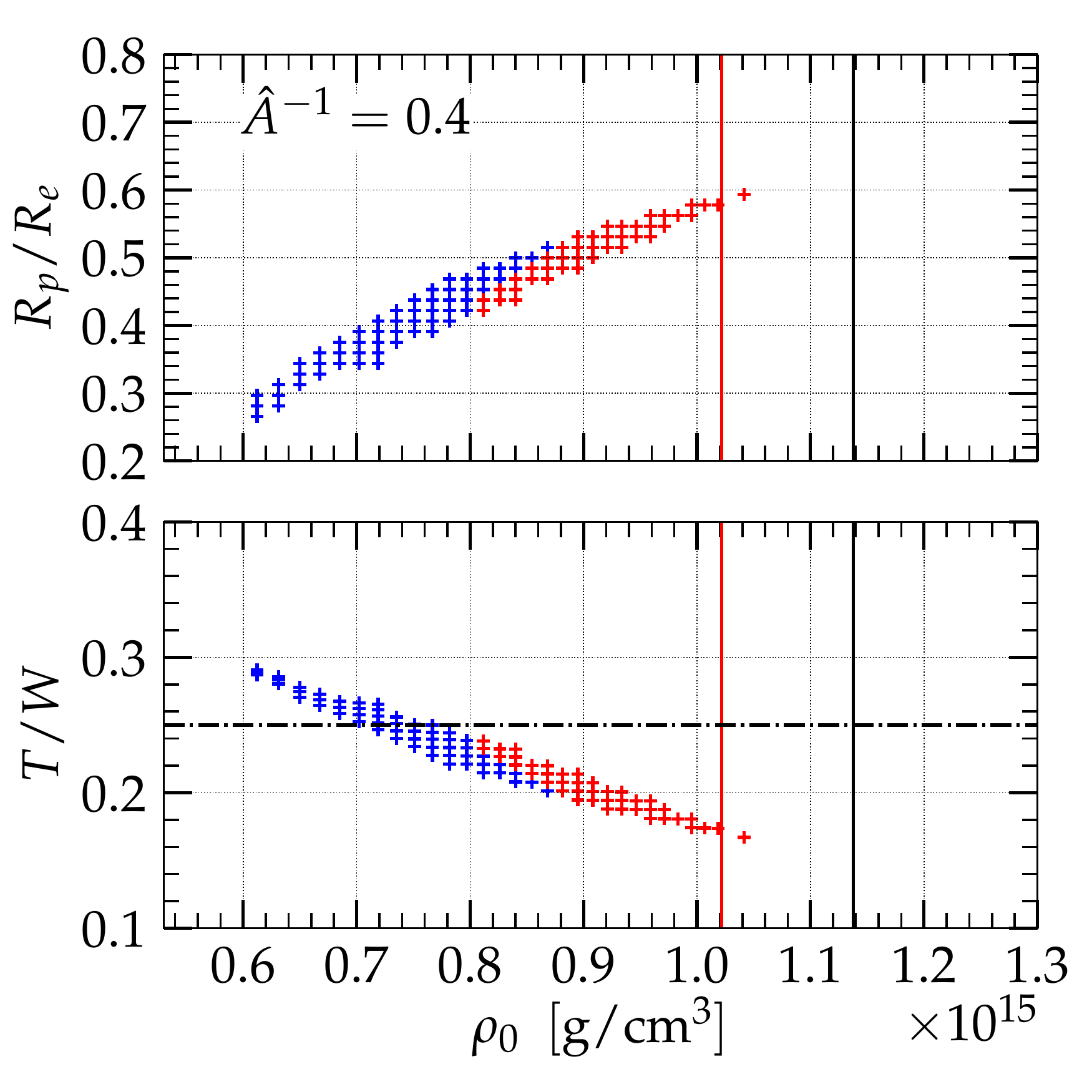}
\includegraphics[width=0.68\columnwidth]{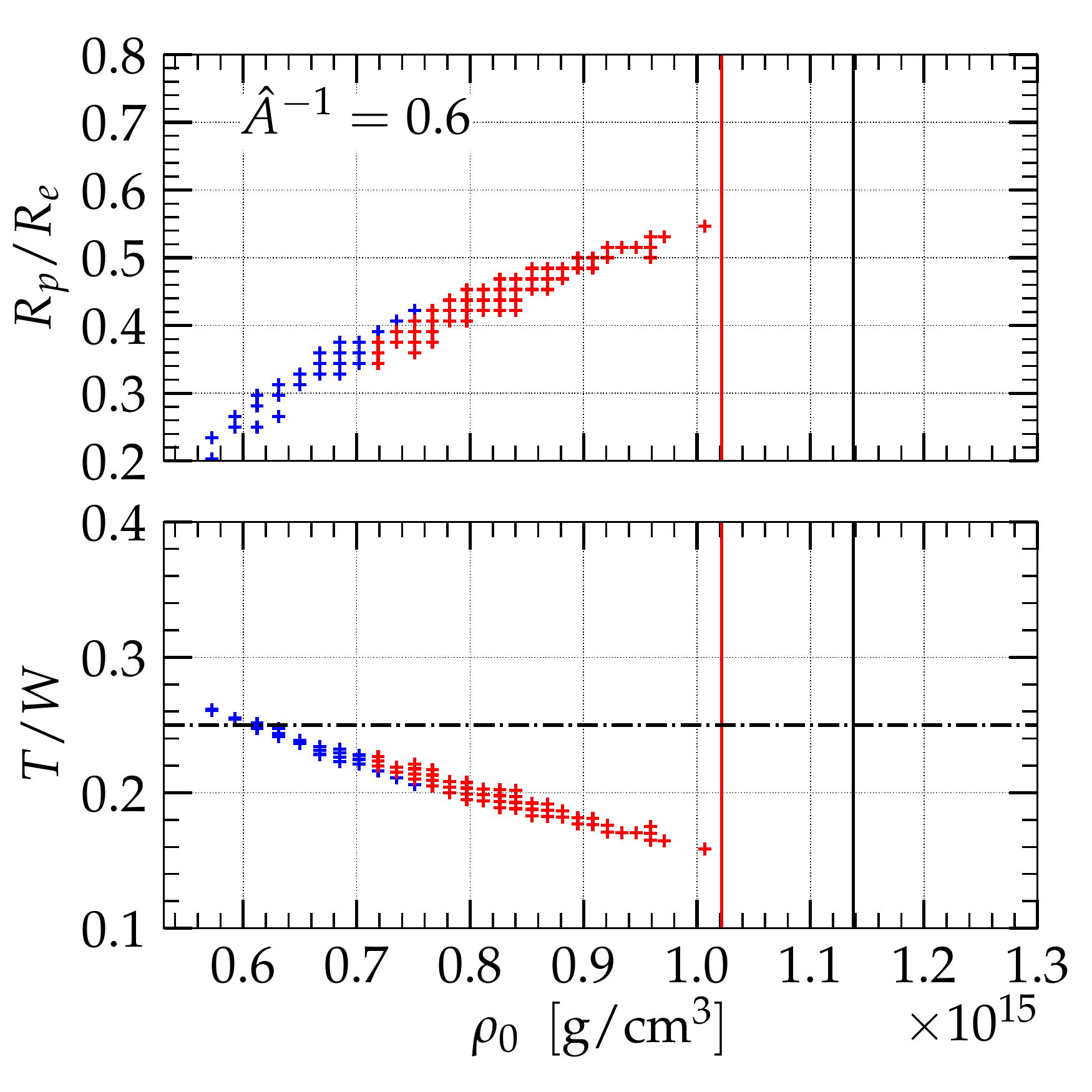}
\includegraphics[width=0.68\columnwidth]{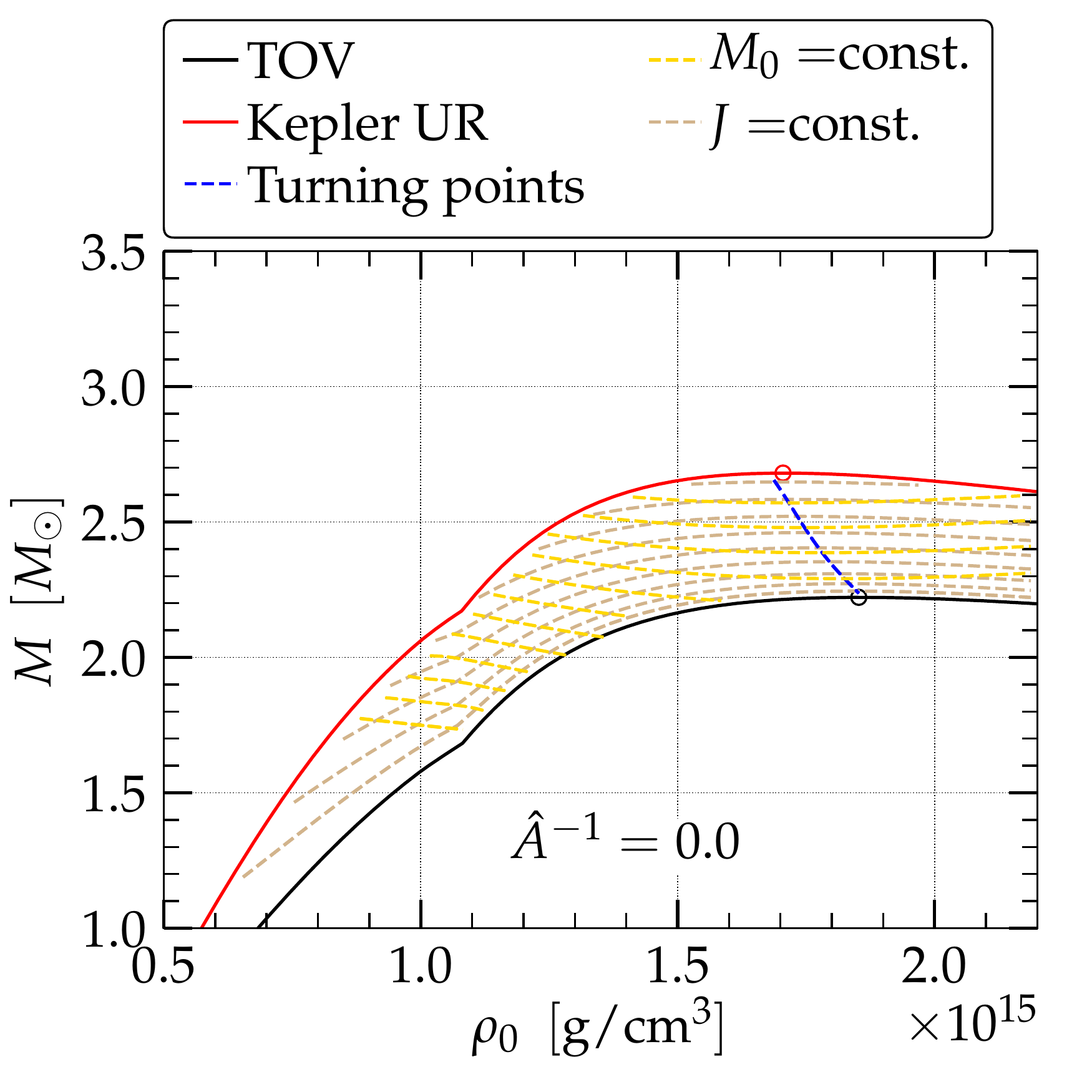}
\includegraphics[width=0.68\columnwidth]{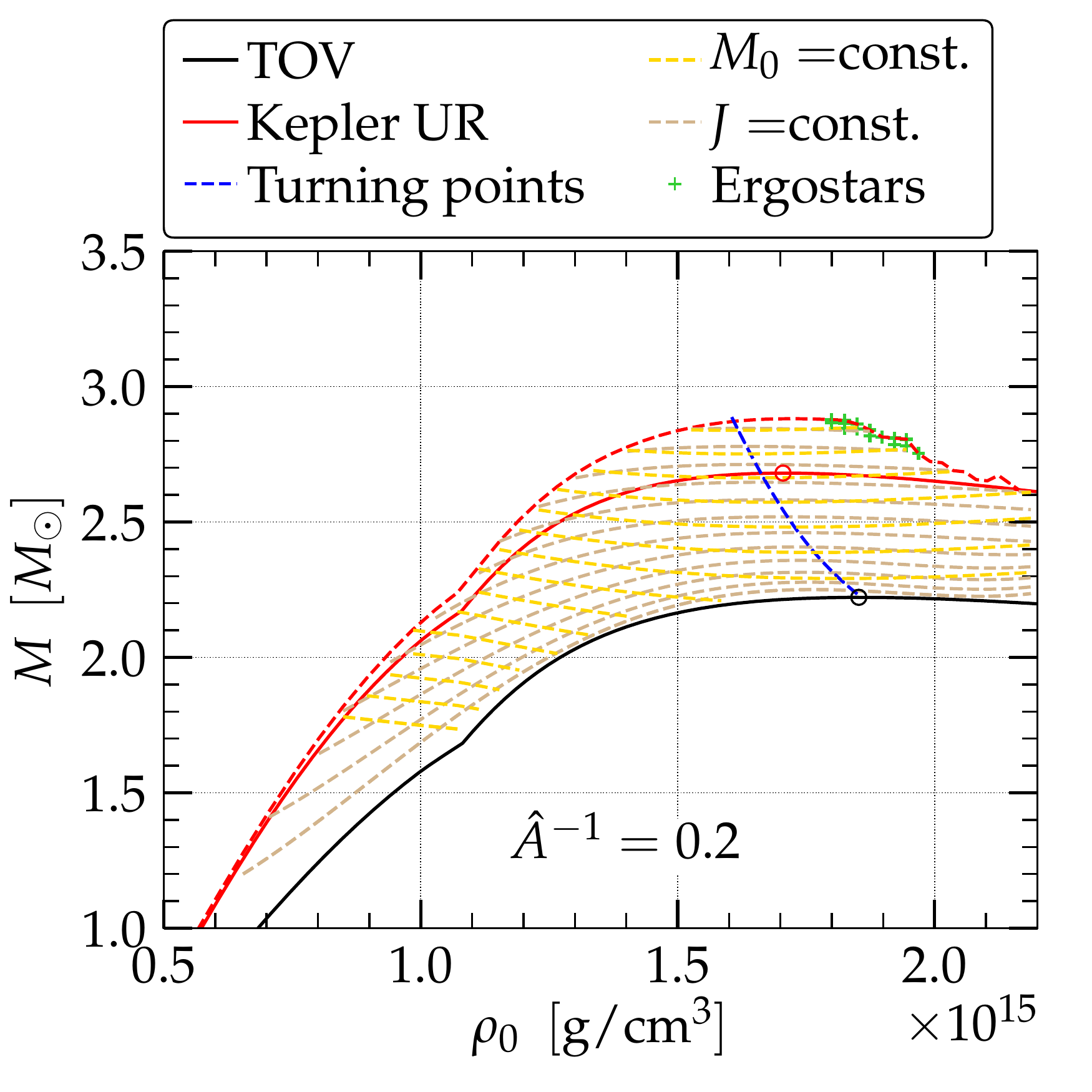}
\includegraphics[width=0.68\columnwidth]{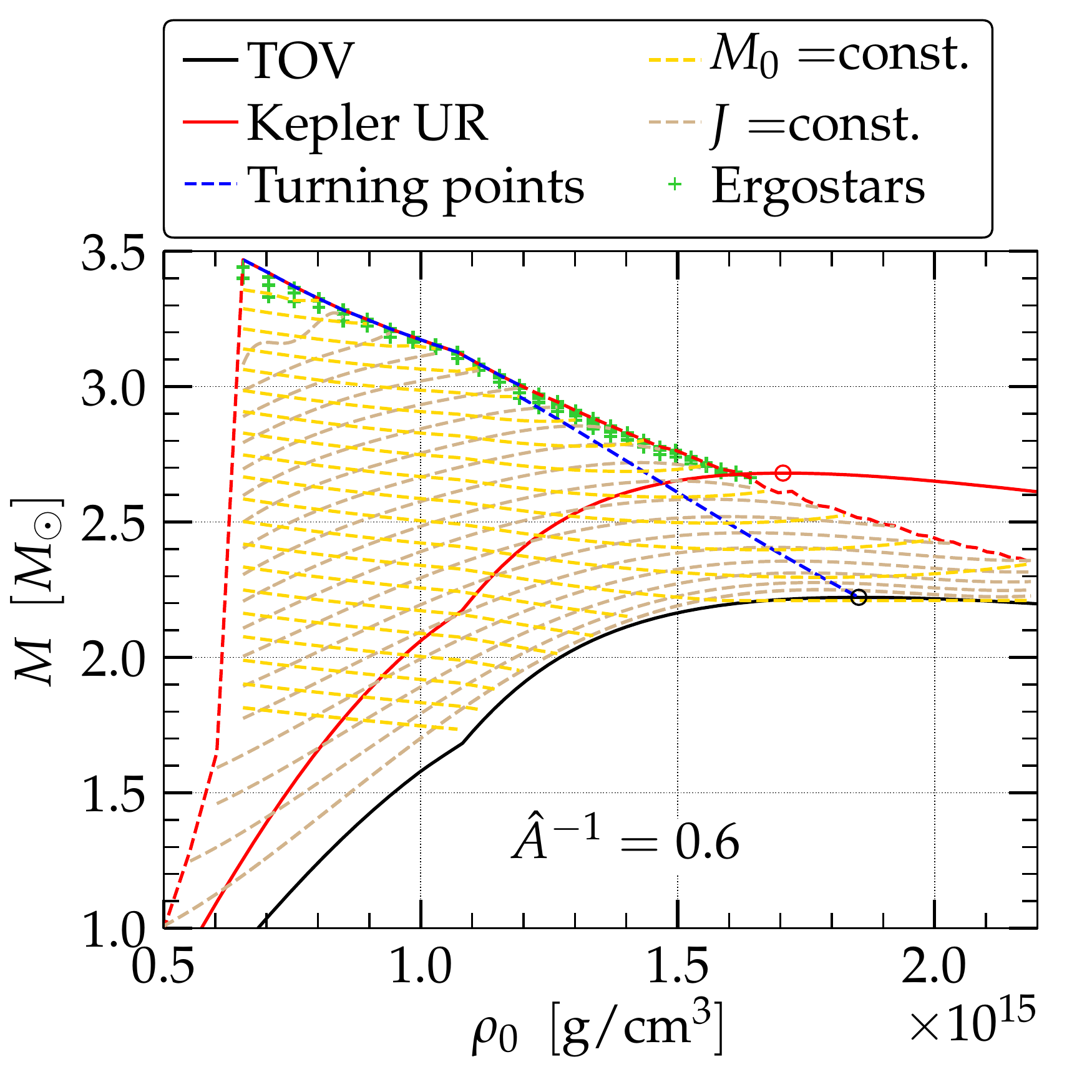}
\includegraphics[width=0.68\columnwidth]{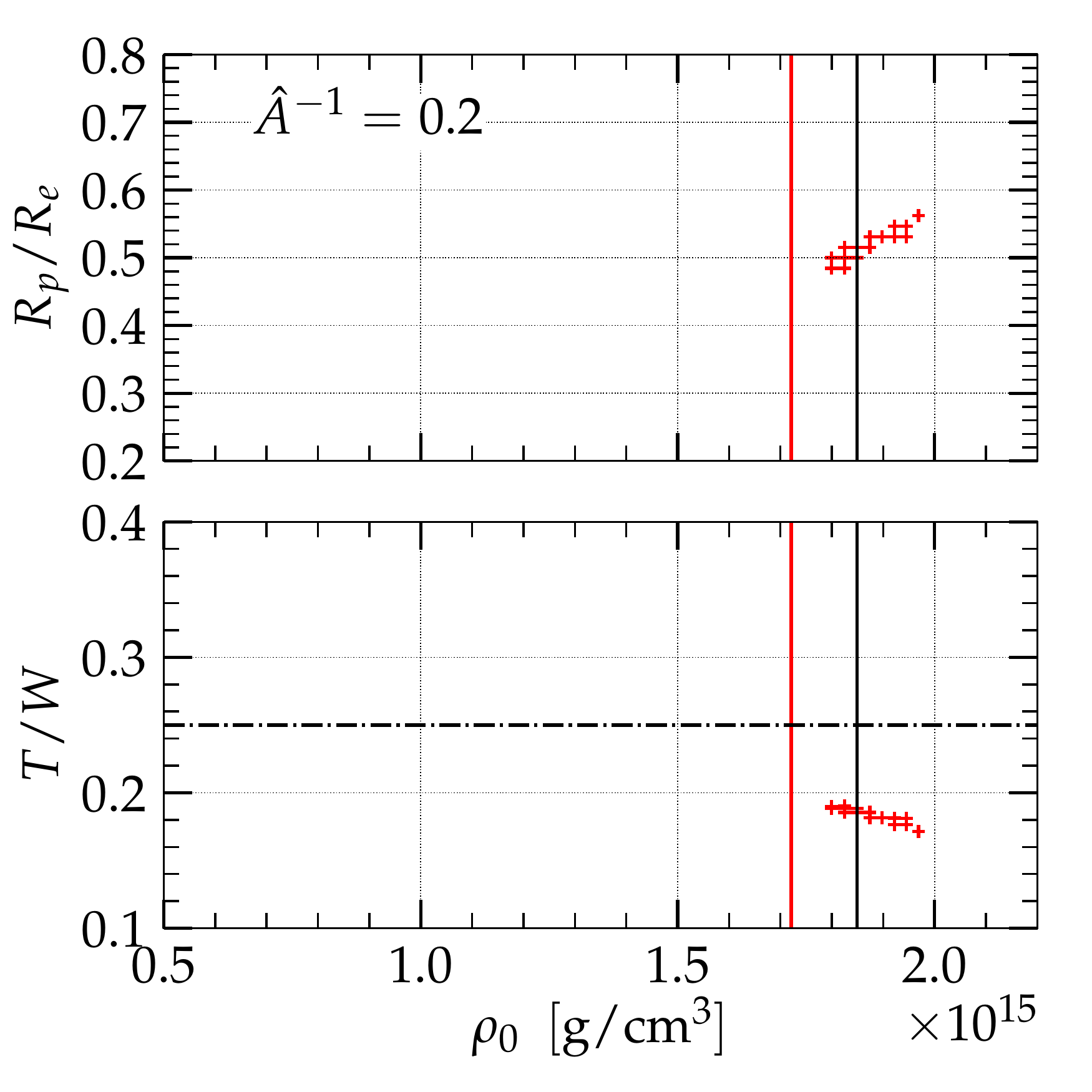}
\includegraphics[width=0.68\columnwidth]{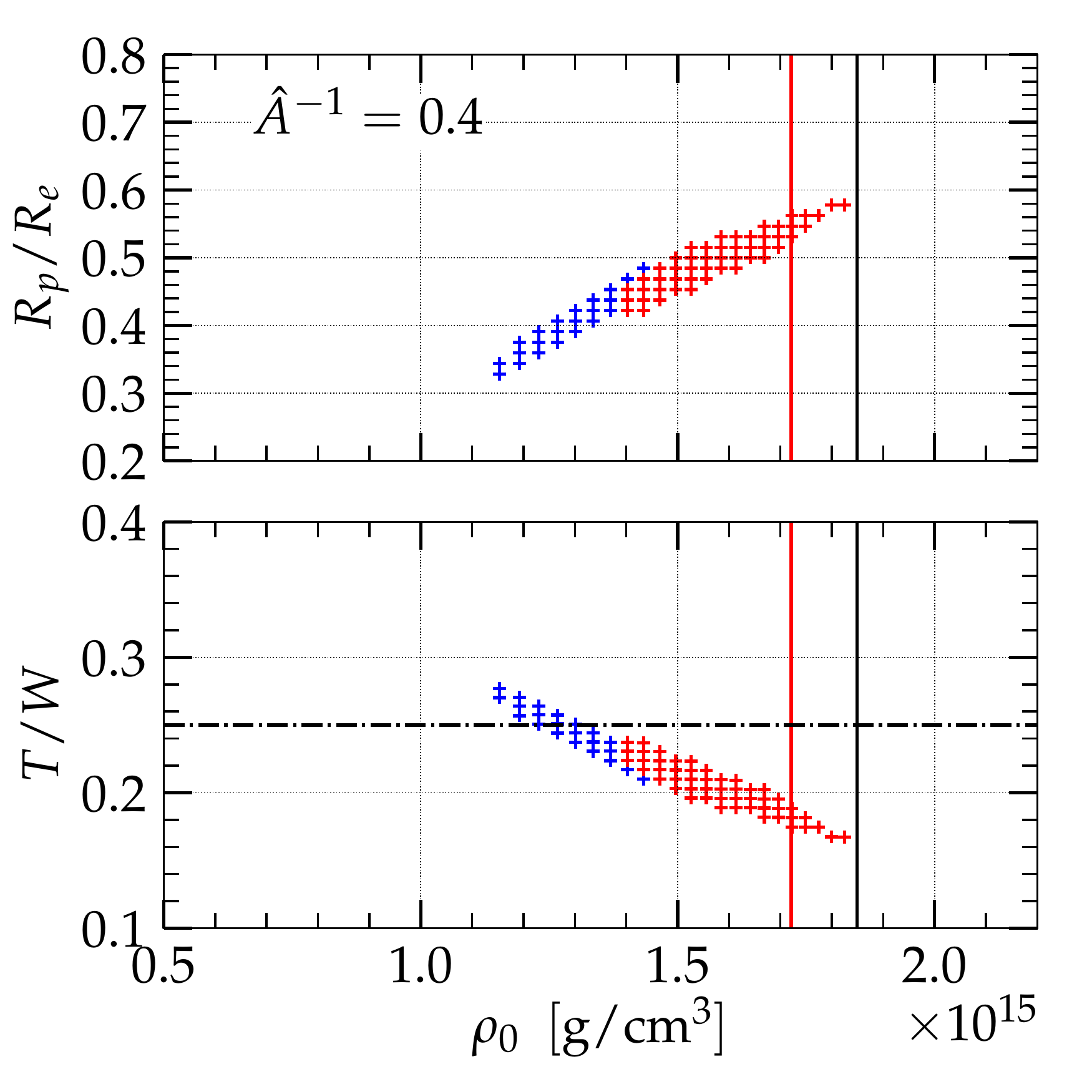}
\includegraphics[width=0.68\columnwidth]{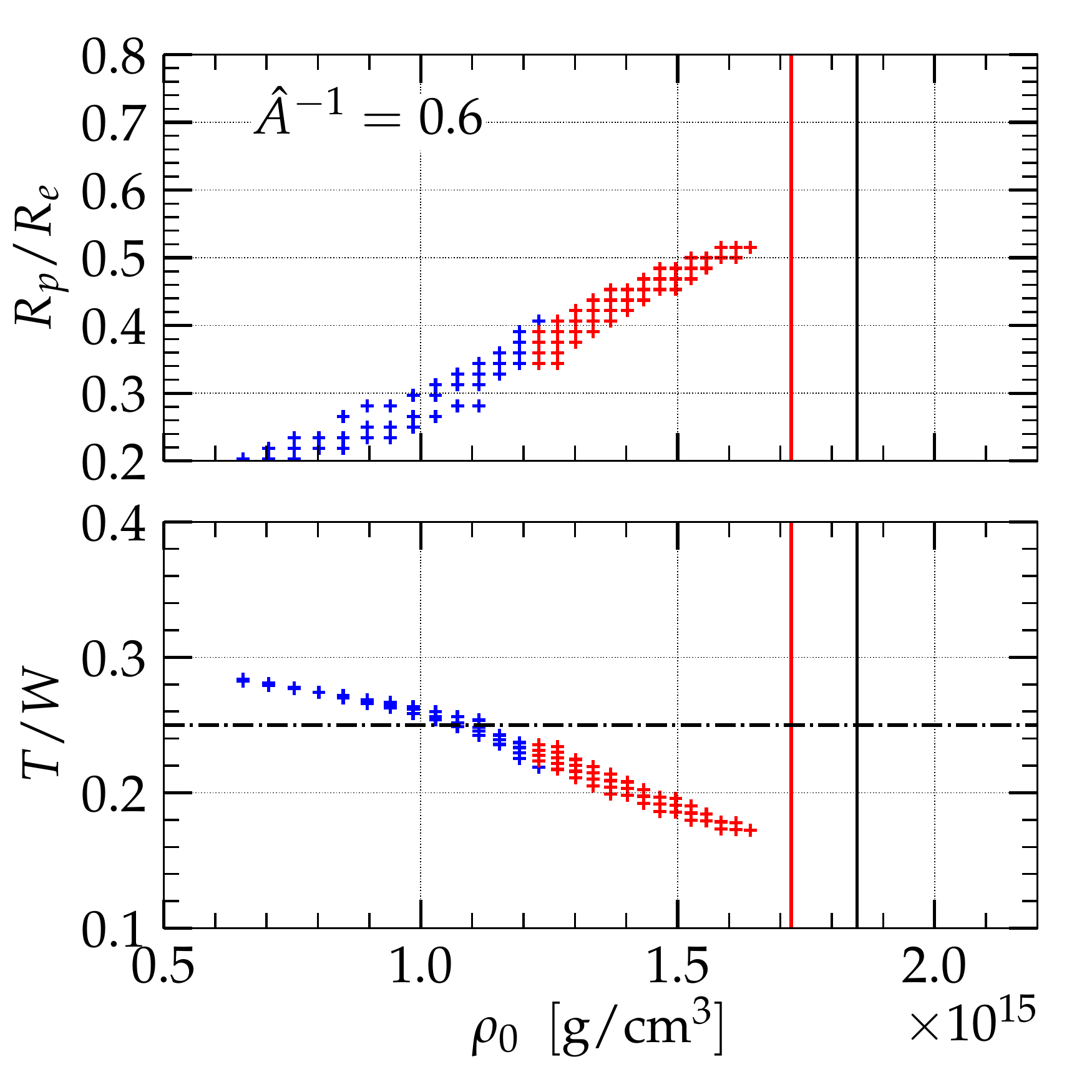}
\caption{Same as Fig. \ref{fig:SLy} but for the SLycc2 (first and second rows) and 
SLycc4 (third and fourth rows) EOSs.}
\label{fig:SLycc24}
\end{center}
\end{figure*}

When differential rotation is considered, the general trends are (1) the turning point line moves
up and turns counterclockwise with respect to the maximum TOV mass point, as in Fig. \ref{fig:SLy}. 
(2) The ergostars move toward smaller densities well
beyond the turning point line, toward the stable part of the parameter space. (3) For a larger degree
of differential rotation, the ergostars tend to accumulate toward the turning point line and also the 
number of them tends to decrease. When differential rotation is large enough, the ergostars almost disappear. 

The fact that a very mild differential rotation moved the ergostars from the unstable regime at high
densities on the right of the turning point line to the left at lower central densities enabled us
to find dynamically stable models \cite{Tsokaros:2019mlz}. Although these models 
used a different EOS (ALF2cc), they do not differ significantly from the models of Fig. \ref{fig:SLycc1}
since apart from a small crust, the rest of the star has the same (causal) EOS. In particular, the featured 
model in Fig. 1 of \cite{Tsokaros:2019mlz} had $\hat{A}^{-1}=0.2$, a central density 
$\GR_0 = 4.52\times 10^{14}\ {\rm g/cm^3}$, and mass $M=5.709\ M_\odot$. Looking at the right panel
of the top row in Fig. \ref{fig:SLycc1}, we can see that indeed such a model lies within the dynamically
stable regime.

From the bottom row panels of Fig. \ref{fig:SLycc1}, where differential rotation is small ($\hat{A}^{-1}=0.1$) 
all ergostars are spheroidal for the SLycc1 EOS, and they progressively become toroidal with higher differential 
rotation (this is in contrast with the SLy EOS where almost all ergostars found had toroidal topology). Also,
$T/W$ for the spheroidal models is below the benchmark value of $0.25$, while it becomes larger and reaches
the $0.3$ value as differential rotation is increased. We note here that for $\hat{A}^{-1}=0.1$, the ergostars
to the right of the blue line (which are the majority of them) should be unstable to axisymmetric perturbations.
For the small number of ergostars to the left of the red line the possibility of dynamical stability is 
significant. For $\hat{A}^{-1}=0.2$, all models with density larger than approximately 
$\GR_0\approx 5\times 10^{14}\ {\rm g/cm^3}$ should also be unstable to axisymmetric perturbations, but
the ones with less central density can be stable even with respect to nonaxisymmetric modes (there are
many models with $T/W<0.25$ and even some with $T/W>0.25$ can be stable). Similar arguments
can be made for $\hat{A}^{-1}=0.4$.

When the causal core is assumed at $2\GR_{0\rm nuc}$, ergostars almost disappear from the uniformly
rotating regime (Fig. \ref{fig:SLycc24} top left panel). Similar to the SLycc1 EOS, small differential 
rotation ($\hat{A}^{-1}=0.2$) brings ergostars into the stable side of the turning point line
and, according to Fig. \ref{fig:SLycc24} second row left panel, these are 
possibly stable against nonaxisymmetric perturbations.
As the degree of differential rotation increases, the turning point line turns counterclockwise with respect to
the maximum spherical mass point, and the ergostars accumulate toward the end point of our convergence
regime. The middle and right panels in the second row of Fig. \ref{fig:SLycc24} show the deformation and $T/W$
when $\hat{A}^{-1}=0.4,0.6$ for the SLycc2 EOS. Full simulations will be needed to probe the fate of these
equilibria. The bottom two rows in Fig. \ref{fig:SLycc24} depict the ergostars when the causal core shifts 
at $4\GR_{0\rm nuc}$. Here, the EOS is very close to the original SLy, apart from the very high density regime;
thus, the position of the ergostars resembles the one found in Fig. \ref{fig:SLy}.

\section{Discussion}

It has recently been proposed \cite{Komissarov:2004ms,2005MNRAS.359..801K} that 
the mechanism behind the launching of relativistic jets from compact objects
is the ergosphere and not a black hole horizon.
In \cite{Ruiz:2012te}, the authors tested a simplified version of this scenario by performing a force-free 
numerical simulation of a homogeneous ergostar using the Cowling approximation. They confirmed that the 
Blandford-Znajek mechanism is not directly related to the horizon of the black hole by showing that
(a) the magnetic field collimation, (b) the induced charged density and poloidal currents, and (c) the
electromagnetic luminosity that are produced by a rotating ergostar are similar to those observed in 
a rotating black hole spacetime.
Their use of an incompressible EOS together with their freezing of the gravitational 
field, raises doubts regarding the stability of ergostars in a realistic evolutionary scenario. 
As in \cite{Ruiz:2018wah,Ruiz:2016rai,prs15}, we define an incipient jet based on the 
following three characteristics: (1) a collimated, tightly wound magnetic field, (2) a mildly relativistic 
outflow ($\Gamma_L>1.2$), and 
(3) the outflow is confined by a funnel containing a (nearly) force-free magnetic field $b^2/(2\GR_0)>1$.
Here, $b^2=B^2/(4\pi)$, and $B$ is the magnetic field at the poles. In the case of \cite{Ruiz:2012te},
the absence of all matter does not permit conditions (2) and (3) to be checked, which motivates our efforts 
to explore the parameter space of dynamically stable ergostars with a compressible and causal EOS.

Regarding the Friedman instability, it was shown that the $m=2$ bar mode of a homogeneous ergostar having a 
period $T=27M$ has a growth time
$\GT_{_{\rm F}}\approx 10^8 M $ \cite{1996MNRAS.282..580Y}. Larger values of $m$ have
even larger $\GT_{_{\rm F}}$. On the other hand, the Alfv\'en timescale is
\be
\frac{\GT_{_{\rm A}}}{M} = \frac{\sqrt{4\pi\GE}}{B} \left(\frac{R}{M}\right) \ .
\label{eq:alfven}
\ee
For typical ergostars
and $B\sim 10^{12}\ {\rm G}$, one gets $\GT_{_{\rm A}}/M\sim 10^6 \sim \GT_{_{\rm F}}/100$.
Therefore, it is improbable that the Friedman instability will have any effect on the possible
formation of a jet. On the other hand, large magnetic fields ($\gtrapprox 10^{15}\ {\rm G}$) are
needed to bring the Alfv\'en timescale on levels that can be currently simulated ($\lessapprox 10^3 M$)
but sufficiently small that they are not dynamically significant initially.

In this work, we constructed more than $\sim 30,000$ uniformly and differentially 
rotating equilibria using four EOSs in order to probe the parameter space and identify the parameters 
under which ergostars appear. The most favorable parameters will be adopted in the future for full 
magnetohydrodynamical simulations. Using the SLy
EOS as a basis, we constructed three other EOSs by imposing a causal core at $\GR_{0\rm nuc},2\GR_{0\rm nuc}$,
and $4\GR_{0\rm nuc}$. We expect that similar behavior will be found when any other EOS is used instead
of the SLy one. The differential rotation law that we explored is the so-called ``$j$-const'' law, and
it will be interesting in the future to see how robust our findings are when other differential rotating
laws are employed, like those presented in \cite{Uryu:2017obi} that model more accurately the rotation
profile of a neutron star merger remnant. 
In all cases considered, we calculated the 
turning point line \cite{1988ApJ...325..722F} and commented on the stability properties of the ergostars
that we found. For a regular
EOS like the SLy, most ergostars appear on the unstable side of the turning point line, but for 
small differential rotation, models on the stable side also appear. These stars typically are highly
hypermassive and very close to the limits of convergence for the CST code. Their stability will have to be
probed by full general relativistic simulations as in \cite{Tsokaros:2019mlz}. For an EOS like SLycc1, ergostars
appear more frequently for a mild degree of differential rotation. Here, dynamically stable models exist, as 
shown in \cite{Tsokaros:2019mlz}. Given the fact that the stars of this EOS resemble quark stars, we conjecture
that stable ergostars of quark or strange matter will have more favorable possibilities for existence. 
When a causal core is found deep in the high density regime of a neutron star, the number of ergostars
that we were able to construct  diminished.

\section{Acknowledgments.}

This work was supported by National Science Foundation Grant No. PHY-1662211 and 
the National Aeronautics and Space Administration (NASA) Grant No. 80NSSC17K0070 to the
University of Illinois at Urbana-Champaign. This work made use of the
Extreme Science and Engineering Discovery Environment, which is supported by National
Science Foundation Grant No. TG-MCA99S008. This research is part of the Blue Waters
sustained-petascale computing project, which is supported by the National Science Foundation
(Grants No. OCI-0725070 and No. ACI-1238993) and the State of Illinois. Blue Waters
is a joint effort of the University of Illinois at Urbana-Champaign and its National Center
for Supercomputing Applications. Resources supporting this work were also provided by the
NASA High-End Computing Program through the NASA Advanced  Supercomputing Division at Ames 
Research Center.   

\bibliographystyle{apsrev4-1}
\bibliography{references}

\end{document}

%% file: kigou.tex
\newcommand{\beq}{\begin{equation}} 
\newcommand{\eeq}{\end{equation}} 
\newcommand{\beqn}{\begin{eqnarray}} 
\newcommand{\eeqn}{\end{eqnarray}}

\newcommand{\zD}{{\raise1.0ex\hbox{${}^{\ \circ}$}}\!\!\!\!\!D}
\newcommand{\alone}{{\raise0.5ex\hbox{${}^{\ 1}$}}\!\!\!\!\alpha}

\newcommand{\nalam}{\mathrel{\raise0.9ex\hbox{$^\lambda$}\mkern-14mu
\lower0.0ex\hbox{$\nabla$}}}

\newcommand{\zeroD}{{\raise1.0ex\hbox{${}^{\ \circ}$}}\!\!\!\!\!D}

\newcommand{\zLap}{{\raise1.0ex\hbox{${}^{\ \circ}$}}\!\!\!\!\Delta}
\newcommand{\zna}{{\raise1.0ex\hbox{${}^{\ \circ}$}}\!\!\!\!\!\nabla}
\newcommand{\zS}{{\raise1.0ex\hbox{${}^{\ \circ}$}}\!\!\!\!\!S}

